\documentclass[12pt,letterpaper]{article}
\usepackage{jcgs-v10}
\usepackage{comment}
\usepackage{mathrsfs}

\newsiamthm{assumption}{Assumption}
\newsiamthm{example}{Example}
\newsiamremark{remark}{Remark}

\crefname{assumption}{assumption}{assumptions}
\Crefname{assumption}{Assumption}{Assumptions}
\crefname{remark}{remark}{remarks}
\Crefname{remark}{Remark}{Remarks}
\crefname{definition}{definition}{definitions}
\Crefname{definition}{Definition}{Definitions}

\crefname{appendix}{appendix}{appendices}
\Crefname{appendix}{Appendix}{Appendices}

\newenvironment{shrinkeq}[1]{\bgroup
	\addtolength\abovedisplayshortskip{#1}
	\addtolength\abovedisplayskip{#1}
	\addtolength\belowdisplayshortskip{#1}
	\addtolength\belowdisplayskip{#1}}
{\egroup\ignorespacesafterend}

\newcommand{\E}{\mathbb E}

\newcommand{\Unif}{\operatorname{Unif}}

\newcommand{\one}{\mathbf 1}

\newcommand{\eps}{\varepsilon}

\newcommand{\dd}{\,\mathrm d}

\headers{Array-RQMC for Metropolis--Hastings}{J. Du, J. Dick, Z. Pan, and H. Zhu}

\title{Fixed-Horizon Error Rates for Array-RQMC Metropolis--Hastings}
\author{%
Jiarui Du%
\thanks{School of Mathematics, South China University of Technology,
Guangzhou 510641, People's Republic of China
(\email{mascut.dujiarui@mail.scut.edu.cn}). Jiarui Du is supported by the China Scholarship Council program (Project ID: 202506150076).}
\and
Josef Dick%
\thanks{School of Mathematics and Statistics,
The University of New South Wales, NSW 2052, Australia
(\email{josef.dick@unsw.edu.au}). Josef Dick is supported by ARC grant DP220101811.}
\and
Zexin Pan%
\thanks{Institute of Fundamental and Transdisciplinary Research,
Zhejiang University, Hangzhou 310058, People's Republic of China
(\email{zep002@zju.edu.cn}).}
\and
Houying Zhu%
\thanks{Corresponding author. School of Mathematical and Physical Sciences, Macquarie
University, NSW 2109, Australia(\mbox{\email{houying.zhu@mq.edu.au}}). Houying Zhu is supported by ARC grant DP260100348.}
}

\ifpdf
\hypersetup{
  pdftitle={Fixed-Horizon Error Rates for Array-RQMC Metropolis--Hastings},
  pdfauthor={Jiarui Du, Josef Dick, Zexin Pan, and Houying Zhu}
}
\fi

\begin{document}
\spacingset{1}
\maketitle

\begin{abstract}
This paper studies randomised quasi-Monte Carlo (RQMC) methods for Metropolis--Hastings (MH) algorithms, whose accept--reject discontinuity complicates RQMC analysis. To circumvent some of these complications, we study Array-RQMC for MH, with $N$ replicated chains evolved over a fixed time horizon $T$, where we make use of Hilbert sorting for joint multidimensional updates at each step. For one-dimensional MH and bounded-variation test functions, we prove a sampling mean squared error bound of $\mathcal O_T(N^{-4/3})$ relative to the finite-horizon expectation. For product targets, coordinatewise Array-RQMC Metropolis-within-Gibbs retains this exponent for additive marginal bounded-variation tests, with constants that may depend on dimension. With Gaussian random-walk proposals, the bounds extend to scalar and additive marginal polynomials with an additional factor $(\log N)^\ell$, where $\ell$ is the maximum polynomial degree. For joint updates in state dimension $p$ using $p$ proposal uniforms, we provide a conditional $\mathcal O_T(N^{-1-1/(p+2)})$ framework under regularity assumptions that are not verified for the joint multidimensional examples.

Numerical experiments support the verified one-dimensional theory and demonstrate variance reduction for joint Hilbert-sorted updates, including Bayesian logistic regression, while illustrating the limitations of these gains as dimension increases.
\end{abstract}



\vfill
\newpage
\spacingset{1.8}

\section{Introduction}

Monte Carlo (MC) is a standard approach for approximating target
expectations, which we denote by \(\Pi(f)=\int_{\mathcal X}f(\bm{x})\pi(\bm{x})\,\dd \bm{x}\), where $\pi$ is a target density on a state space $\mathcal X$ and $f$
is a test function. In many applications, particularly in Bayesian
computation, direct sampling from the target distribution $\Pi$ is
infeasible and $\pi$ is known only up to a normalizing constant. Markov chain Monte
Carlo (MCMC), and in particular the Metropolis--Hastings (MH) algorithm,
addresses this problem by constructing a Markov chain with invariant distribution $\Pi$ and
estimating $\Pi(f)$ from its sample path; see, for example,
\citet{robert2004}. In standard Monte Carlo integration, the driving
inputs are independent and identically distributed (IID), whereas
quasi-Monte Carlo (QMC) replaces them by more evenly distributed deterministic point
sets with the aim of obtaining more accurate numerical estimates.
Randomized QMC (RQMC) randomizes such point sets while preserving their
uniformity properties and permits error assessment through independent
replications \citep{bookdick2010}. This naturally raises
the question of whether QMC uniformity can also improve the
accuracy of practical MH algorithms.

One approach is Markov chain quasi-Monte Carlo (MCQMC), initiated by
\citet{owen2005}, which replaces the successive IID inputs
along one chain trajectory by a completely uniformly distributed (CUD) or
weakly CUD  sequence. Numerical studies
\citep{harase2021,du2025unbiased} have
reported substantial variance reductions and, for several Gibbs
samplers, empirical convergence rates faster than Monte Carlo. For MH,
however, the reported gains are often more
modest. The difficulty is that low-discrepancy structure must survive a
long nonlinear composition with an acceptance discontinuity at
every transition. Available MCQMC theory establishes consistency under
suitable conditions \citep{tribble2008,chen2011}, while {{convergence-rate}}
results rely on existence arguments \citep{dick2014,dick2016}; in particular, they do not
provide rates for the explicit CUD constructions commonly used in
practice.

Array-RQMC instead imposes QMC uniformity across a particle population at
each fixed transition. The randomized construction was introduced in
\citet{lecuyer2008}; mapping alternatives and convergence behavior were
studied further in \citet{lecuyer2009}, and the method, sorting
strategies, and known convergence results were reviewed in
\citet{lecuyer2018}. It has since been applied in a range of simulation
settings, including option pricing, stochastic reaction networks, and, more recently,
walk-on-spheres simulation
\citep{benabdellah2019,puchhammer2021,ho2026walk}.
It evolves \(N\) particles in parallel: at each time, it sorts the
population, matches the particles to a freshly randomized low-discrepancy
point set, and advances them simultaneously. Its fresh
randomization at each transition prevents the nominal RQMC dimension from
growing with $T$. For MH, this population structure, which is local in time,
turns the acceptance discontinuity into an indicator integration problem of
fixed dimension at each step. Existing convergence theory provides two useful
benchmarks. {{\citet{gerber2015}}} established an
$o(N^{-1})$ variance result for randomized sequential QMC under regularity
conditions, with corresponding Array-RQMC conclusions obtained through
Hilbert ordering of multidimensional states. Earlier,
{{\citet{lecuyer2008}}} proved the sharper explicit rate
$\mathcal O(N^{-3/2})$ for one-dimensional monotone inversion transitions
under a square-stratified sampling construction and additional regularity
assumptions. These results do not directly provide a rate for the usual
proposal-and-accept/reject implementation of MH, which we analyze
through the geometry of local indicator events.

Because Array-RQMC is intrinsically a population method, its natural
MC benchmark is $N$ independent copies of the same chain
evolved over the same finite horizon. For a fixed horizon $T$, averaging these
copies has MSE \(\mathcal{O}_T(N^{-1})\) for the corresponding
expectation under a finite second moment. The question addressed here is
whether the dependence deliberately introduced by Array-RQMC can
improve this rate in $N$ for a standard MH transition.

Our analysis concerns sampling error relative to the finite-horizon
expectation $\mu_T(f)$ with $T$ fixed. The Array-RQMC estimator is
unbiased for this expectation; unless the chain starts in stationarity,
the estimation of $\Pi(f)$ may contain an additional bias independent of $N$.
In one dimension, explicit variation conditions give
$\mathcal O_T(N^{-4/3})$ for 
bounded-variation test functions. For product targets, coordinatewise
Array-RQMC Metropolis-within-Gibbs (MWG) retains this exponent for additive
marginal bounded-variation tests. This is a dimension-independent exponent, but not a bound with dimension-uniform constants. With Gaussian proposals,
an extension to estimating the expectation of polynomials has the additional factor $(\log N)^\ell$ for maximum polynomial degree $\ell$.
For joint $p$-dimensional MH, suitable Hardy--Krause variation
conditions and uniform boundary-tube bounds yield a conditional $\mathcal O_T(N^{-1-1/(p+2)})$ bound. These hypotheses do not follow from Hilbert sorting and are not verified for the
joint examples. 

Numerical experiments support the verified rates and
contrast this product construction with joint Hilbert-sorted MH in
increasing dimension.

\Cref{sec:background} reviews Array-RQMC and Hilbert ordering;
\Cref{sec:main} gives the MH algorithm, the multidimensional geometric result
and the verifiable one-dimensional and product-target theory.
\Cref{sec:numerics} reports
experiments, and \Cref{sec:conclusion} discusses the remaining questions.
The appendices prove the multidimensional and one-dimensional fixed-horizon
theorems. The supplementary materials follow the one-dimensional argument: 
supporting lemmas, verification tools and target classes, the product-target
MWG reduction and polynomial-test extension, the scrambled-net inversion
extension and horizon dependence. MATLAB scripts (MATLAB R2024b) are available from \url{https://github.com/Jiarui-Du/Array-RQMC-MH}.

\section{Background and Array-RQMC}
\label{sec:background}

\subsection{Notation}
\label{sec:notation}
For a joint MH update in state dimension \(p\), we consider proposals
generated from \(p\) uniform variables. $D$ denotes the dimension of
each Array-RQMC point set: $D=p+2$ for joint MH, including the acceptance
and sorting/matching coordinates, and $D=3$ for each coordinatewise
MWG update in \Cref{sec:product-mwg}. Vectors and vector-valued maps
are written in bold, for example $\bm{x}=(x_1,\ldots,x_p)$; scalar
components and one-dimensional state variables are written in ordinary type.
We write $\lambda_D$ for the $D$-dimensional Lebesgue measure and
$\one_E$ for the indicator of a set $E$.
{{For a real random variable $Z$, write $\|Z\|_2:=(\mathbb E|Z|^2)^{1/2}$, and for a
sigma-field $\mathcal G$, write
$\|Z\|_{2\mid\mathcal G}:=(\mathbb E[|Z|^2\mid\mathcal G])^{1/2}$.
In our results this expectation is $\mathbb E_{\mathrm{RQMC}}$, taken over the
point-set randomizations (\Cref{sec:finite-horizon}). For a deterministic
function $g$ on $[0,1)^D$, $\|g\|_2$ denotes its $L^2(\lambda_D)$ norm.}}

For a real-valued function \(h\) on an interval \(\mathcal X \subseteq\mathbb R\), let
\begin{shrinkeq}{0em}
\[
V_{\mathcal X}(h):=
\sup_{\substack{r\ge1\\x_0<\cdots<x_r,\;x_\ell\in\mathcal X}}
\sum_{\ell=1}^r |h(x_\ell)-h(x_{\ell-1})|,
\]
\end{shrinkeq}
and write \({\rm BV}(\mathcal X)\) for the class of bounded variation functions satisfying
\(V_{\mathcal X}(h)<\infty\). Constants hidden in \(\mathcal{O}_T(\cdot)\)
depend on the fixed horizon $T$, but not on \(N\).

\subsection{Array-RQMC and Hilbert ordering}
\label{sec:array-rqmc}
Array-RQMC evolves $N$ copies of a Markov chain
$\bm{X}_j=\bm{\varphi}_j(\bm{X}_{j-1},\bm{\Theta}_j)$ using a fresh
randomized low-discrepancy point set at each transition
\citep{lecuyer2008,lecuyer2009,lecuyer2018}, where
$\bm{\varphi}_j:\mathcal X\times[0,1)^{D-1}\to\mathcal X$ and the
ordinary chain has independent $\bm{\Theta}_j\sim\Unif([0,1)^{D-1})$.

Suppose first that a scalar sorting key $\mathfrak h:\mathcal X\to\mathbb R$
has been chosen. At step $j$, order the old particles by this key,
writing the resulting permutation as $\varsigma_j$, so that
$\mathfrak h(\bm{X}_{\varsigma_j(0),j-1})\le\cdots\le
\mathfrak h(\bm{X}_{\varsigma_j(N-1),j-1})$.
Generate a freshly randomized point set
$P_{N,j}=\{(V_{k,j},\bm{\Theta}_{k,j})\}_{k=0}^{N-1}\subset[0,1)^D$.
The first coordinate $V_{k,j}$ serves only for sorting and matching;
the remaining coordinates $\bm{\Theta}_{k,j}\in[0,1)^{D-1}$ drive the
transition. Sort the complete points by their first coordinate,
$V_{(0),j}\le\cdots\le V_{(N-1),j}$, retaining each point's associated
transition inputs, and match the particles and points by rank:
\[
 \bm{X}_{i,j}=\bm{\varphi}_j
 (\bm{X}_{\varsigma_j(i),j-1},\bm{\Theta}_{(i),j}),
 \qquad i=0,\ldots,N-1.
\]
Fresh randomization is used at every transition, independently across
time, while within a transition the low-discrepancy structure couples
ordered particle ranks with transition inputs. The aim is to approximate
the corresponding Markov-chain distribution more accurately than $N$
independently updated particles; in one dimension this can be viewed as
improving the empirical CDF approximation.

The step function
\begin{equation}\label{eq:hilbert-step}
 \bm{Z}_{N,j-1}(v)=\bm{X}_{\varsigma_j(i),j-1},
 \qquad v\in[i/N,(i+1)/N),\quad i=0,\ldots,N-1,
\end{equation}
records the old particle assigned to each rank interval. When the first
coordinate is exactly $N$-stratified,
$\bm{Z}_{N,j-1}(V_{(i),j})=\bm{X}_{\varsigma_j(i),j-1}$ almost surely,
so evaluating the update through this step function gives precisely the
rank-matched particle update. The scrambled-net condition used here
is stated in \Cref{ass:points}.

The construction therefore requires an ordering of the particle states.
For $p=1$, we use the natural numerical order, $\mathfrak h(x)=x$. 
For joint updates with $p>1$, there is no canonical natural ordering. Among the batch,
split and Hilbert-based sorts reviewed in \citet[Section~3]{lecuyer2018},
we use Hilbert ordering following \citet{gerber2015}.
Choose a componentwise continuous strictly increasing transformation
\(
 \bm{\psi}(\bm{x})
 =
 (\psi_1(x_1),\ldots,\psi_p(x_p)):
 \mathcal X\to[0,1]^p,
\)
and write
\(
 \mathcal J:=\bm{\psi}(\mathcal X)\subseteq[0,1]^p.
\)
Thus $\bm{\psi}$ is a bijection from $\mathcal X$ onto
$\mathcal J$, with componentwise inverse
$\bm{\psi}^{-1}:\mathcal J\to\mathcal X$.
Componentwise logistic or Gaussian-CDF transformations on unbounded
state spaces are examples
\citep[Section~2.4]{gerber2015}{{.}} Let
$\bm{\mathscr H}_p:[0,1]\to[0,1]^p$ be a Hilbert curve and choose a
Borel measurable right inverse $\iota_p:[0,1]^p\to[0,1]$ satisfying
$\bm{\mathscr H}_p(\iota_p(\bm{z}))=\bm{z}$
\citep[Proposition~2]{gerber2019resampling}. The scalar sorting key is
$\mathfrak h(\bm{x})=\iota_p(\bm{\psi}(\bm{x}))$.
In computation we use a finite-resolution Hilbert index
\citep[Section~3.2]{lecuyer2018}. The transformation $\bm{\psi}$ is used only for sorting; 
the target, proposal and MH update remain defined on the original state space.

\Needspace{7\baselineskip}
\section{Array-RQMC MH and fixed-horizon results}
\label{sec:main}

\subsection{The Array-RQMC-MH algorithm}

Let $\mathcal X\subseteq\mathbb R^p$ be a nonempty Borel state space.
All densities and update maps are Borel measurable, all density values
used in acceptance ratios are finite, and kernels and their parameters
are deterministic. Let $\Pi$ have density $\pi$ on $\mathcal X$, and fix
$\bm{x}_0\in\{\bm{z}\in\mathcal X:\pi(\bm{z})>0\}$. At step $j$, the proposal density
is $q_j(\cdot\mid \bm{z})$ and a measurable generator $\bm{Y}_j(\bm{z},\bm{u})$,
$\bm{u}\in[0,1)^p$, produces this proposal for uniform $\bm{u}$.
Values on null endpoint sets can be completed by $\bm{Y}_j(\bm{z},\bm{u})=\bm{z}$.
The Metropolis--Hastings acceptance probability is defined by
\begin{shrinkeq}{0em}
\begin{equation}
  \label{eq:alpha}
\alpha_j(\bm{z},\bm{y}):=
\begin{cases}
\displaystyle
1\wedge
\dfrac{\pi(\bm{y})q_j(\bm{z}\mid \bm{y})}{\pi(\bm{z})q_j(\bm{y}\mid \bm{z})},
& \pi(\bm{z})q_j(\bm{y}\mid \bm{z})>0,\\[1mm]
0,
& \pi(\bm{z})q_j(\bm{y}\mid \bm{z})=0{{,}}
\end{cases}
\end{equation}
\end{shrinkeq}
where $1\wedge x = \min\{1,x\}$.
Set $a_j(\bm{z},\bm{u})=\alpha_j(\bm{z},\bm{Y}_j(\bm{z},\bm{u}))$. With $w$ the acceptance
uniform, the update on $\mathcal X\times[0,1)^p\times[0,1)$ is
\begin{equation}\label{eq:mh-update-general}
 \bm{\Psi}_j(\bm{z},\bm{u},w)= \bm{Y}_j(\bm{z},\bm{u})\one\{w\le a_j(\bm{z},\bm{u})\}
 +\bm{z}\one\{w>a_j(\bm{z},\bm{u})\}.
\end{equation}
The full transition input is $(\bm{u},w)$; in the Array-RQMC construction, $\bm{U}$ always denotes
the $p$ proposal uniforms and $\Omega$ denotes the acceptance uniform.

A $(t,m,D)$-net in base $b$, with $N=b^m$, places exactly $b^t$
points in every base-$b$ elementary interval of volume $b^{t-m}$
{{\citep{owenqmc2023}}}. Nested uniform scrambling preserves the net property
almost surely and makes each point marginally uniform on $[0,1)^D$
\citep{owen1995randomly,bookdick2010}.

\begin{assumption}[Point-set condition]
\label[assumption]{ass:points}
Fix a base \(b\geq2\) and a quality parameter \(t\geq0\), both
independent of \(m\). For \(N=b^m\), \(m\ge t\), and every \(j=1,\ldots,T\), let
\begin{shrinkeq}{0em}
\[
 P_{N,j}=\{\bm{W}_{k,j}:k=0,\ldots,N-1\},\qquad
 \bm{W}_{k,j}:=(V_{k,j},\bm{U}_{k,j},\Omega_{k,j})\in[0,1)^D.
\]
\end{shrinkeq}
Here $V_{k,j}\in[0,1)$ is the sorting coordinate,
$\bm{U}_{k,j}\in[0,1)^p$ supplies the proposal, and
$\Omega_{k,j}\in[0,1)$ supplies the acceptance decision.
Assume $P_{N,j}$ is a nested uniform scramble of a $(t,m,D)$-net in base $b$.
The scrambles are mutually independent across
\(j=1,\ldots,T\). In addition, almost surely,
\begin{shrinkeq}{0em}
\begin{equation}\label{eq:first-strat}
     \#
  \left\{
    k:
    V_{k,j}
    \in
    \left[
      \frac{i}{N},
      \frac{i+1}{N}
    \right)
  \right\}
  =
  1,
  \qquad
  i=0,\ldots,N-1.
\end{equation}
\end{shrinkeq}
\end{assumption}

\begin{remark}\label[remark]{rem:points}
    A general \((t,m,D)\)-net with \(t>0\) need not have an exactly
    \(N\)-stratified one-dimensional projection. For the
    first \(2^m\) points of a $D$-dimensional Sobol' sequence, however,
    the first coordinate projection is a \((0,m,1)\)-net in base \(2\) \citep{owenqmc2023}.
    It therefore contains exactly one point in each interval $\left[
      i/2^m,(i+1)/2^m
    \right)$, $i=0,\ldots,2^m-1$.
Nested uniform scrambling preserves this projection property almost
surely. Hence a nested uniform scramble of the first \(2^m\) points of
such a Sobol' sequence satisfies \Cref{ass:points}.
\end{remark}

With point sets satisfying \Cref{ass:points}, initialize
\(\bm{X}_{i,0}=\bm{x}_0\) for \(i=0,\ldots,N-1\).
Using the Array-RQMC construction of \Cref{sec:array-rqmc} with the MH
map \eqref{eq:mh-update-general}, update
\begin{equation}\label{eq:array-update}
 \bm{X}_{i,j}=\bm{\Psi}_j(\bm{X}_{\varsigma_j(i),j-1},\bm{U}_{(i),j},\Omega_{(i),j}),
 \qquad i=0,\ldots,N-1.
\end{equation}
For \(0\le T_0<T\), the resulting estimator is
\begin{shrinkeq}
{0em}
\begin{equation}
\label{eq:array-estimator}
\widehat\mu_{N,T}(f)
:=
\frac{1}{N(T-T_0)}
\sum_{j=T_0+1}^{T}
\sum_{i=0}^{N-1} f(\bm{X}_{i,j}).
\end{equation}
\end{shrinkeq}

\subsection{Finite-horizon target and stationary bias}
\label{sec:finite-horizon}
For the fixed initial state $\bm{x}_0$, let $\bm{X}_0=\bm{x}_0$ and
$\bm{X}_j=\bm{\Psi}_j(\bm{X}_{j-1},\bm{U}_j,\Omega_j)$ with independent uniform
$(\bm{U}_j,\Omega_j)\in[0,1)^p\times[0,1)$. For Borel $A\subseteq\mathcal X$, set
\[
 \mathsf P_j(\bm{z},A)=\int_{[0,1)^p}\int_0^1
       \one\{\bm{\Psi}_j(\bm{z},\bm{u},w)\in A\}\,\dd w\,\dd\bm{u}.
\]
{{For measurable $h:\mathcal X\to\mathbb R$, bounded or nonnegative, and a
probability measure $\nu$ on $\mathcal X$, write
\[
\begin{aligned}
 \mathsf P_jh(\bm{z})&:=\int_{\mathcal X}h(\bm{y})\,\mathsf P_j(\bm{z},\dd\bm{y})
 =\int_{[0,1)^p}\int_0^1 h\bigl(\bm{\Psi}_j(\bm{z},\bm{u},w)\bigr)\,\dd w\,\dd\bm{u},\\
 (\nu\mathsf P_j)(A)&:=\int_{\mathcal X}\mathsf P_j(\bm{z},A)\,\nu(\dd\bm{z}),
\end{aligned}
\]
so that $\mathsf P_j\one_A(\bm{z})=\mathsf P_j(\bm{z},A)$ and
$\int h\,\dd(\nu\mathsf P_j)=\int\mathsf P_jh\,\dd\nu$.}}
Each kernel leaves $\Pi$ invariant. Let $\nu_j$ be the law of $\bm{X}_j$,
so that $\nu_0=\delta_{\bm{x}_0}$ and $\nu_j=\nu_{j-1}\mathsf P_j$.
Here $\delta_{\bm{y}}$ denotes the point mass at $\bm{y}$. The empirical measure is
$\widehat\nu_{N,j}:=N^{-1}\sum_{i=0}^{N-1}\delta_{\bm{X}_{i,j}}$.
Expectations and variances over the point-set randomizations are denoted
by $\mathbb E_{\mathrm{RQMC}}$ and $\operatorname{Var}_{\mathrm{RQMC}}$.
For a test function \(f\in L^1(\nu_j)\), write
\begin{shrinkeq}{0em}
\[
\nu_j(f):=\int_{\mathcal X}f\,\dd\nu_j,
\qquad
\widehat\nu_{N,j}(f):=\frac1N\sum_{i=0}^{N-1}f(\bm{X}_{i,j}).
\]
\end{shrinkeq}
Exact {{first-coordinate}} stratification and independence of the randomizations
across time give {{(see \Cref{app:innovation-identities})}}, for every \(j\),
\begin{shrinkeq}{0em}
\begin{equation}\label{eq:one-time-unbias}
\mathbb E_{\mathrm{RQMC}}[\widehat\nu_{N,j}(f)]=\nu_j(f).
\end{equation}
\end{shrinkeq}
This is the analogue of {{\citet[Proposition~1]{lecuyer2008}}}. Let \(L:=T-T_0\), and define
\begin{shrinkeq}{0em}
\begin{equation}
\label{eq:est}
\mu_T(f):=\frac1L\sum_{j=T_0+1}^{T}\nu_j(f),\qquad
{{\text{so that}}}\qquad\widehat\mu_{N,T}(f)\mathrel{{{=}}}\frac1L\sum_{j=T_0+1}^{T}\widehat\nu_{N,j}(f).
\end{equation}
\end{shrinkeq}
Consequently, $\widehat\mu_{N,T}(f)$ is an unbiased estimator of \(\mu_T(f)\).
Whenever the relevant second moments are finite,
\begin{shrinkeq}{0em}
\begin{equation}\label{eq:target}
\mathbb E_{\mathrm{RQMC}}[(\widehat\mu_{N,T}(f)-\Pi(f))^2]
=\mathbb E_{\mathrm{RQMC}}[(\widehat\mu_{N,T}(f)-\mu_T(f))^2]
 +(\mu_T(f)-\Pi(f))^2.
\end{equation}
\end{shrinkeq}
The first term is the Array-RQMC sampling error at the fixed horizon,
whereas the second term is the squared bias at the finite horizon and is
independent of \(N\).
If the kernel is homogeneous{{, $\mathsf P_j\equiv\mathsf P$,}} and geometrically ergodic from \(\bm{x}_0\), with
\(\|\mathsf P^j(\bm{x}_0,\cdot)-\Pi\|_{\mathrm{TV}}\le M(\bm{x}_0)\rho^j\), for some \(M(\bm{x}_0)<\infty\) and \(\rho\in(0,1)\) \citep{robertsRosenthal2004}{{, where the total-variation distance is the supremum over Borel sets}}, then for every
bounded measurable $f${{, with $C_f(\bm{x}_0):=2\|f\|_\infty M(\bm{x}_0)$}},
\begin{shrinkeq}{0em}
\[
|\mu_T(f)-\Pi(f)| =\left|\frac1L\sum_{j=T_0+1}^{T}
    \{\mathsf P^jf(\bm{x}_0)-\Pi(f)\}\right|
\le \frac{C_f(\bm{x}_0)}L\sum_{j=T_0+1}^{T}\rho^j
\le \frac{C_f(\bm{x}_0)\rho^{T_0+1}}{L(1-\rho)}.
\]
\end{shrinkeq}
Thus the bias at the finite horizon is classically controllable under geometric ergodicity.
We therefore analyze the first term in
\eqref{eq:target} with \(T\) fixed.

\Needspace{8\baselineskip}
\subsection{Multidimensional fixed-horizon theorem}
\label{sec:md-geometric}
For the multidimensional result, let $\mathcal X=\mathcal X_1\times\cdots\times\mathcal X_p$
be a product of intervals, possibly with infinite endpoints, and use the
componentwise increasing sorting transform $\bm{\psi}$ of \Cref{sec:background}.
Let $\mathcal F_{N,j}$ be the sigma-field generated by the point-set
randomizations through time $j$, with $\mathcal F_{N,0}$ trivial.
For $\bm{x}\in\mathbb R^p$, define the lower orthant
$A_{\bm{x}}:=\mathcal X\cap\prod_{r=1}^p(-\infty,x_r]$ and its local event
\begin{equation}\label{eq:md-local-event}
 B_{N,s,\bm{x}}:=\{(v,\bm{u},w)\in[0,1)^D:
       \bm{\Psi}_s(\bm{Z}_{N,s-1}(v),\bm{u},w)\in A_{\bm{x}}\}.
\end{equation}
For $B\subset[0,1)^D$, let
$(\partial B)_\varepsilon:=\{\bm{y}\in\mathbb R^D:
\operatorname{dist}_2(\bm{y},\partial B)\le\varepsilon\}$,
where $\partial B$ denotes the boundary in $\mathbb R^D$. For a deterministic bounded Borel test $f$, define
\begin{equation}\label{eq:md-propagated}
 G_s:=\sum_{j=\max\{s,T_0+1\}}^T\mathsf P_{s+1:j}f,
 \qquad s=1,\ldots,T.
\end{equation}
Here $\mathsf P_{a:b}=\mathsf P_a\cdots\mathsf P_b$ acts on functions,
with an empty product equal to the identity. Each $G_s$ is deterministic
and bounded. Write $V_{\rm HK}$ for the usual Hardy--Krause variation
anchored at $\bm{1}${{; see \citet{aistleitner2015functions}}} for details.

\begin{theorem}[Multidimensional fixed-horizon bound]
\label{thm:md-fixed}
For Array-RQMC MH on the product state space above, suppose that the
point sets satisfy \Cref{ass:points} and that all particles are
initialized at the same deterministic state $\bm{x}_0$.
Fix integers $0\le T_0<T$, put $L=T-T_0$, and let
$f:\mathcal X\to\mathbb R$ be a deterministic bounded Borel test.
For each $s=1,\ldots,T$, assume that
$G_s\circ\bm{\psi}^{-1}:\mathcal J\to\mathbb R$
admits a deterministic coordinatewise left-continuous extension
$\widetilde G_s:[0,1]^p\to\mathbb R$
with
\(
 V_{\rm HK}(\widetilde G_s)<\infty.
\)
Also suppose that there exist deterministic constants $0\le M_s<\infty$
and $\varepsilon_{0,s}>0$, independent of the test function, such that
\begin{equation}\label{eq:md-orthant-tube}
 \lambda_D((\partial B_{N,s,\bm{x}})_\varepsilon)\le M_s\varepsilon,
 \qquad 0<\varepsilon<\varepsilon_{0,s},
\end{equation}
uniformly over admissible $N$, $\bm{x}\in\mathbb R^p$, and realized old
particle arrays. Then there exists $C_{b,t,D}<\infty$, depending only
on $b,t,D$, such that, for all sufficiently large admissible $N$,
\begin{equation}\label{eq:md-fixed-bound}
 \mathbb E_{\mathrm{RQMC}}[(\widehat\mu_{N,T}(f)-\mu_T(f))^2]
 \le\frac{C_{b,t,D}}{L^2}
       \sum_{s=1}^T M_s V_{\rm HK}(\widetilde G_s)^2\,N^{-1-1/D}.
\end{equation}
\end{theorem}
\begin{proof}
See \Cref{app:md-proof}.
\end{proof}
Since \(D=p+2\), for fixed \(p\) and \(T\) the bound gives sampling MSE \(\mathcal O_T(N^{-1-1/(p+2)})\){{, provided \eqref{eq:md-orthant-tube} holds.}}

\subsection{One-dimensional specialization}
\label{sec:general-main}
For this subsection and \Cref{sec:verification}, let $p=1$, $D=3$,
and let $\mathcal X\subseteq\mathbb R$ be a nonempty interval, possibly
unbounded. The lower orthants in \Cref{thm:md-fixed} reduce to half-lines
$A_x:=\mathcal X\cap(-\infty,x]$, $x\in\mathbb R$.
We use the natural ordering of the particles and an inverse-transform
proposal with distribution function ${{F_j}}(\cdot\mid z)$:
\begin{shrinkeq}{0em}
\begin{equation}\label{eq:proposal}
    Y_j(z,u)
  :=
  \inf\{y\in\mathcal X:{{F_j}}(y\mid z)\ge u\},
  \qquad u\in(0,1).
\end{equation}
\end{shrinkeq}
The scalar update $\Psi_j$ is given by \eqref{eq:mh-update-general},
with the endpoint convention $Y_j(z,0)=z$, so that
$\Psi_j(z,0,w)=z$ for every $w\in[0,1)$. For $x\in\mathbb R$ and $z\in\mathcal X$, define the transition CDF
$K_j(x\mid z):=\mathsf P_j(z,A_x)$.

Fix $1\le j\le T$. For $x\in\mathbb R$ and $z\in\mathcal X$, set
$\beta_{j,x}(z):={{F_j}}(x\mid z)=\int_{A_x}q_j(y\mid z)\,\dd y$.
For $z\in\mathcal X$ and $0<u<1$, set
$a_j(z,u):=\alpha_j(z,Y_j(z,u))$.
For any $N\ge1$, write the sorted states as $z_i:={{X_{\varsigma_j(i),j-1}}}$,
with $z_0\le\cdots\le z_{N-1}$, and use the scalar version $Z_{N,j-1}$
of \eqref{eq:hilbert-step}. Set
$a_{N,j}(v,u)=a_j(Z_{N,j-1}(v),u)$.
For a bounded function $g$ on a rectangle $\mathcal R$, write
\begin{equation}\label{eq:osc}
 \operatorname{osc}_{\mathcal R}(g):=\sup_{\bm{y},\bm{y}'\in\mathcal R}|g(\bm{y})-g(\bm{y}')|.
\end{equation}
Use the grid intervals
$J_\jmath ^{(k)}=[\jmath /k,(\jmath +1)/k]\cap[0,1)$ and
$K_q^{(k)}=[q/k,(q+1)/k]\cap(0,1)$, where $k\ge1$ and
$\jmath ,q=0,\ldots,k-1$. Define
\begin{equation}\label{eq:general-quantities}
\begin{aligned}
 C_{a,j}&:=\sup_{\substack{N\ge1,\;z_i\in\mathcal X\\z_0\le\cdots\le z_{N-1}}}
          \sup_{k\ge1}\frac1k\sum_{\jmath ,q=0}^{k-1}
          \operatorname{osc}_{J_\jmath ^{(k)}\times K_q^{(k)}}(a_{N,j}),\\
 C_{\beta,j}&:=\sup_{x\in\mathbb R}V_{\mathcal X}(\beta_{j,x}),
 \qquad \Lambda_j:=\sup_{x\in\mathbb R}
 V_{\mathcal X}(K_j(x\mid\cdot)).
\end{aligned}
\end{equation}
The following two lemmas connect these quantities to the two ingredients
of \Cref{thm:md-fixed}.

In one dimension, Hardy--Krause variation reduces to ordinary total
variation, and natural sorting requires no transformation of the state
space. We therefore control the propagated tests directly through
$V_{\mathcal X}(G_s)$; the pointwise-BV transfer below replaces the
transformed HK condition in \Cref{thm:md-fixed} without requiring left
continuity. 
\begin{lemma}[Pointwise BV transfer and propagation]
\label{lem:bv-transfer}
Let $\Delta$ be the difference of two random probability measures on an interval
$\mathcal X$. If, for every real $x$,
\(
 \|\Delta(A_x)\|_2\le {C},
\)
then for every pointwise
$f\in\mathrm{BV}(\mathcal X)$,
$\|\Delta(f)\|_2\le {C} V_{\mathcal X}(f).$ The same conclusion holds for conditional
$L^2$ norms when the half-line bounds hold conditionally with a deterministic
bound ${C}$. Also, a Markov kernel $\mathsf P$ satisfying
$\sup_{x\in\mathbb R} V_{\mathcal X}(K(x\mid\cdot))\le\Lambda$ obeys
\begin{equation}\label{eq:kernel-bv-operator}
 V_{\mathcal X}(\mathsf P f)\le\Lambda V_{\mathcal X}(f),
\end{equation}
where $K(x\mid z):=\mathsf P(z,A_x)$ {{ and $\mathsf P f$ is defined as for
$\mathsf P_j$ in \Cref{sec:finite-horizon}}}.
\end{lemma}

\begin{proof}
The proof uses a Jordan decomposition and half-line level sets; see
\Cref{sec:supp-bv-transfer} of the supplementary materials.
\end{proof}

For fixed $0\le T_0<T$, pointwise $f\in\mathrm{BV}(\mathcal X)$ and
finite $\Lambda_j$, define 
\begin{equation}\label{eq:horizon-weight}
 \mathcal H_s^{T_0,T}:=
 \sum_{j=\max\{s,T_0+1\}}^T\prod_{q=s+1}^j\Lambda_q,
 \qquad s=1,\ldots,T,
\end{equation}
with empty products equal to one. Iterating \eqref{eq:kernel-bv-operator}
and applying the triangle inequality to the sum defining $G_s$ gives
\begin{equation}\label{eq:Gs-bv-bound}
 V_{\mathcal X}(G_s)\le\mathcal H_s^{T_0,T}V_{\mathcal X}(f).
\end{equation}

The scalar specialization of \eqref{eq:md-local-event} is
\begin{equation}\label{eq:event}
 B_{N,j,x}:=\{(v,u,w)\in[0,1)^3:
     \Psi_j(Z_{N,j-1}(v),u,w)\in A_x\}.
\end{equation}

\begin{lemma}[Uniform MH boundary-tube bound]
\label{thm:tube}
If $C_{a,j},C_{\beta,j}<\infty$, then
\begin{equation}\label{eq:B-tube-bound}
 \lambda_3((\partial B_{N,j,x})_\varepsilon)\le M_j\varepsilon,
 \qquad M_j:=14C_{a,j}+6C_{\beta,j}+84,\quad 0<\varepsilon<1/2.
\end{equation}
The bound holds uniformly in $N$, the sorted old array and $x\in\mathbb R$.
\end{lemma}
\begin{proof}
The proof decomposes $B_{N,j,x}$ into accepted-proposal and
rejection regions and bounds the resulting pieces separately;
see \Cref{sec:supp-boundary} of the supplementary materials.
\end{proof}

Together, these lemmas provide the geometric bound and the direct
pointwise-BV transfer used in the following one-dimensional result.

\Needspace{7\baselineskip}
\begin{theorem}[One-dimensional fixed-horizon bound]
\label{thm:main}
Suppose that \Cref{ass:points} holds with $p=1$, $D=3$ for the
one-dimensional Array-RQMC-MH construction above, and that all particles
start from the same deterministic $x_0\in\{z\in\mathcal X:\pi(z)>0\}$.
Assume
\[
 C_{a,j}<\infty,\qquad C_{\beta,j}<\infty,\qquad
 \Lambda_j<\infty,\qquad j=1,\ldots,T.
\]
For fixed $0\le T_0<T$, let $L=T-T_0$ and let
$f\in\mathrm{BV}(\mathcal X)$ be pointwise defined.
Then, for all admissible {{$N\ge N_0$, with $N_0$ given in \eqref{eq:main-cutoff}}},
\begin{equation}\label{eq:main-rate}
 \mathbb E_{\mathrm{RQMC}}
 \left[(\widehat\mu_{N,T}(f)-\mu_T(f))^2\right]
 \le C_{T_0,T}(f)N^{-4/3}, \quad C_{T_0,T}(f) := \frac{{{c_{b,t,3}}}V_{\mathcal X}(f)^2}{L^2}
       \sum_{s=1}^T M_s\bigl(\mathcal H_s^{T_0,T}\bigr)^2{{,}}
\end{equation}
where {{$c_{b,t,3}<\infty$, given in \eqref{eq:tube-variance-constant},}} depends only on $b,t$, and $M_s$ is defined in
\eqref{eq:B-tube-bound}. In particular, the sampling MSE is
$\mathcal O_T(N^{-4/3})$.
\end{theorem}
\begin{proof}
  See \Cref{app:onedim}.
\end{proof}

For a one-uniform inversion transition, $D = 2$, giving $N^{-1-1/2}=N^{-3/2}$;
\Cref{app:inversion} of the supplementary materials gives the corresponding scrambled-net extension
of \citet{lecuyer2008}.

\begin{remark}[Horizon dependence and BV contraction]
\label[remark]{rem:horizon-growth}
The horizon dependence of \eqref{eq:main-rate} is governed
by the local geometry
coefficients and the weights $\mathcal H_s^{T_0,T}$; finite
$\Lambda_j$ alone allows these weights to grow rapidly with $T$.
If there are constants $\bar C_a,\bar C_\beta<\infty$ and $0\le\rho<1$,
independent of the horizon, such that
$C_{a,j}\le\bar C_a$, $C_{\beta,j}\le\bar C_\beta$ and
$\Lambda_j\le\rho$ for every $j$, then
\[
 \mathbb E_{\mathrm{RQMC}}
 [(\widehat\mu_{N,T}(f)-\mu_T(f))^2]
 ={{\mathcal O}}(N^{-4/3}L^{-1}),
\]
for sufficiently large admissible $N$. Indeed,
\(
 \sum_{s=1}^T(\mathcal H_s^{T_0,T})^2
 \le (1-\rho)^{-2}\left\{L+\frac{\rho^2}{1-\rho^2}\right\}.
\)
Substitution into \eqref{eq:main-rate} gives the stated rate.
This consequence requires contraction of the actual
transition-CDF variations. {{\Cref{prop:no-global-bv-contraction}}}
of the supplementary materials rules out such global BV contraction
for homogeneous location random-walk MH on $\mathbb R$. Thus the
fixed-horizon theorem does not supply a joint $(N,T)$ rate for those
chains. 

This obstruction concerns the proof technique rather than the actual
horizon behavior. In the supplementary horizon experiment {{(\Cref{subsec:horizon})}} for the
asymmetric Gaussian-mixture target and the CDF test, the observed
variance is
close to the $L^{-1}$ reference rate, while the $N^{4/3}$-scaled
variance curves remain closely aligned. Thus the empirical behavior is
consistent with an $N^{-4/3}L^{-1}$ scaling even though the global BV
contraction condition fails; proving such a joint rate would require
sharper local or distribution-dependent control.
\end{remark}

\subsection{Checkable verification routes}
\label{sec:verification}
We first give a general bound for transition-CDF variation, then verify
the three quantities in \Cref{thm:main} for symmetric location proposals.

\begin{proposition}[Transition-CDF variation]
\label{prop:kernel-variation}
For an MH kernel on a state interval $\mathcal X$, suppose that
$C_{\beta,j}<\infty$ and
$\int_0^1V_{\mathcal X}(a_j(\cdot,u))\,\dd u<\infty$.
Then
\begin{equation}\label{eq:global-Lambda}
 \Lambda_j\le1+C_{\beta,j}
 +\int_0^1V_{\mathcal X}(a_j(\cdot,u))\,\dd u.
\end{equation}
\end{proposition}
\begin{proof}
See \Cref{app:kernel} of the supplementary materials.
\end{proof}

\begin{proposition}[Symmetric location proposals]
\label{prop:symmetric-location}
Let $\mathcal X=\mathbb R$, and let $\pi:\mathbb R\to(0,\infty)$ be a
continuous probability density. Suppose there is a finite partition
$-\infty=c_0<c_1<\cdots<c_R<c_{R+1}=\infty$ such that $\pi$ is monotone
on every $(c_\ell,c_{\ell+1})$, and suppose
\begin{equation}\label{eq:bpi}
 B_\pi:=\sup_{\xi\in\mathbb R}V_{\mathbb R}
 \left(1\wedge\frac{\pi(\cdot+\xi)}{\pi(\cdot)}\right)<\infty.
\end{equation}
Let $h:\mathbb R\to(0,\infty)$ be a measurable, finite-valued symmetric
probability density, with $h(w)=h(-w)$ for every $w$ and a continuous
strictly increasing CDF $H$. For deterministic scales $\sigma_j>0$, use
\begin{equation}\label{eq:symmetric-proposal}
 q_j(y\mid z)=\sigma_j^{-1}h((y-z)/\sigma_j),\qquad
 Y_j(z,u)=z+\sigma_jH^{-1}(u),\quad 0<u<1.
\end{equation}
Then, for every $j=1,\ldots,T$,
\[
 C_{a,j}\le B_\pi+4R+2,\qquad C_{\beta,j}=1,\qquad
 \Lambda_j\le B_\pi+2.
\]
Hence the three quantitative conditions of \Cref{thm:main} hold.
With its point-set and initialization conditions, the fixed-horizon
sampling MSE is $\mathcal O_T(N^{-4/3})$ for every pointwise
$f\in\mathrm{BV}(\mathbb R)$.
\end{proposition}
\begin{proof}
See \Cref{sec:supp-symmetric} of the supplementary materials.
\end{proof}

\Cref{tab:state-variation} summarizes target classes covered by
\Cref{prop:symmetric-location}; \Cref{app:targets} of the supplementary materials proves the bounds.
Gaussian, Laplace and Student-$t$ proposal increments all satisfy
\eqref{eq:symmetric-proposal}; no proposal moment assumption is required
for the BV theorem.

\begin{table}[htbp]
\centering
\caption{Choices of $R$, bounds on $B_\pi$, and the resulting bounds for the three quantities in \Cref{thm:main}, for representative targets under the symmetric location proposals of \Cref{prop:symmetric-location}.}
\label{tab:state-variation}
\setlength{\tabcolsep}{4pt}
\small
\begin{tabular}{@{}lccccc@{}}
\toprule
Target class & $R$ & $B_\pi$ & $C_{a,j}$ & $C_{\beta,j}$ & $\Lambda_j$\\
\midrule
Positive log-concave & $1$ & $\le1$ & $\le7$ & $1$ & $\le3$\\
Student-$t_\nu$, $\nu>0$ & $1$ & $2$ & $\le8$ & $1$ & $\le4$\\
Double well & $3$ & $\le3$ & $\le17$ & $1$ & $\le5$\\
$J$-component Gaussian mixture & $R_J$ & $\le K_J+1$ & $\le K_J+4R_J+3$ & $1$ & $\le K_J+3$\\
\bottomrule
\end{tabular}
\end{table}
\FloatBarrier

For each fixed $J$, the finite integers $R_J,K_J$ depend only on $J$.
The mixture bounds are uniform over positive weights and precisions,
real centers and positive proposal scales; \Cref{sec:supp-gaussian-mixture}
gives the proof and the resulting uniform fixed-horizon error bound.

\subsection{Product targets and coordinatewise Array-RQMC MWG}
\label{sec:product-mwg}
Let $\Pi_p=\bigotimes_{r=1}^p\Pi_r$ have density
$\pi(\bm{x})=\prod_{r=1}^p\pi_r(x_r)$ on
$\mathcal X=\mathcal X_1\times\cdots\times\mathcal X_p$, with $\Pi_r$ supported on $\mathcal X_r$.
Consider systematic-scan Metropolis-within-Gibbs (MWG). In sweep $j$,
coordinates $r=1,\ldots,p$ are updated in that order, using deterministic
proposals $q_{r,j}(y\mid\bm{x})=q_{r,j}(y\mid x_r)$ generated by the
scalar inverse transform $Y_{r,j}(x_r,u)$. On the positive-density support, 
the coordinate acceptance probability is
\begin{equation}\label{eq:mwg-acceptance}
\begin{aligned}
 \alpha_{r,j}(x_r,y_r;\bm{x}_{-r})
=1\wedge\frac{\pi(y_r,\bm{x}_{-r})q_{r,j}(x_r\mid y_r)}
                    {\pi(x_r,\bm{x}_{-r})q_{r,j}(y_r\mid x_r)}=1\wedge\frac{\pi_r(y_r)q_{r,j}(x_r\mid y_r)}
                    {\pi_r(x_r)q_{r,j}(y_r\mid x_r)}
 =:\alpha_{r,j}(x_r,y_r),
\end{aligned}
\end{equation}
where $\bm{x}_{-r}$ denotes the vector of all coordinates except $r$.
Let $\Psi_{r,j}(x,u,w)$ be the scalar MH map
\eqref{eq:mh-update-general}: it returns $Y_{r,j}(x,u)$ when
$w\le\alpha_{r,j}(x,Y_{r,j}(x,u))$, and $x$ otherwise.

For $i=0,\ldots,N-1$, set
$\bm{X}_{i,j}^{(0)}=\bm{X}_{i,j-1}$.
For $r=1,\ldots,p$, let $\bm{X}_{i,j}^{(r)}$
denote the $i$th full particle vector after the
$r$th coordinate update in sweep $j$, including
the reordering induced by sorting and rank matching. At coordinate $r$, let $\varsigma_{r,j}$ order the
vectors $\bm{X}_{i,j}^{(r-1)}$ by their $r$th coordinate.
Generate a fresh randomized point set
\[
 P_{N,r,j}
 :=\{(V_{k,r,j},U_{k,r,j},\Omega_{k,r,j}):
       0\le k<N\}\subset[0,1)^3.
\]
Sort the points by $V$, retaining the associated
transition inputs, and denote the sorted proposal
and acceptance inputs by
$U_{(i),r,j}$ and $\Omega_{(i),r,j}$.
The coordinate update is
\begin{equation}\label{eq:mwg-coordinate-update}
 X_{i,j,\ell}^{(r)}=
 \begin{cases}
 \Psi_{r,j}\!\left(
 X_{\varsigma_{r,j}(i),j,r}^{(r-1)},
 U_{(i),r,j},\Omega_{(i),r,j}
 \right), & \ell=r,\\[1mm]
 X_{\varsigma_{r,j}(i),j,\ell}^{(r-1)},
 & \ell\ne r,
 \end{cases}{{.}}
\end{equation}
After all $p$ coordinate updates, set
$\bm{X}_{i,j}=\bm{X}_{i,j}^{(p)}$.
For every coordinate update, the point set satisfies
\Cref{ass:points} with $D=3$ and common fixed $b,t$;
the randomizations are independent across coordinates and sweeps.
Here $T,T_0$ count complete sweeps: \eqref{eq:array-estimator} uses
sweep-endpoint particles $\bm{X}_{i,j}$, and $\mu_T(f)$ is the matching
finite-horizon expectation for IID-driven systematic-scan MWG.

\begin{corollary}[Product targets and additive BV tests]
\label{cor:product-mwg}
Initialize all particles at the same deterministic $\bm{x}_0$, with
$\pi_r(x_{0,r})>0$ for every $r$, and suppose each coordinatewise one-dimensional MH construction satisfies the hypotheses of
\Cref{thm:main}. Fix integers $0\le T_0<T$ and let
$f(\bm{x})=\sum_{r=1}^p f_r(x_r)$, with
pointwise $f_r\in\mathrm{BV}(\mathcal X_r)$. Then, for all sufficiently
large admissible $N$,
\begin{equation}\label{eq:product-mwg-rate}
 \mathbb E_{\mathrm{RQMC}}
 [(\widehat\mu_{N,T}(f)-\mu_T(f))^2]
 \le N^{-4/3}\sum_{r=1}^p C_{r,T_0,T}(f_r),
\end{equation}
where $C_{r,T_0,T}(f_r)<\infty$ is supplied by the corresponding scalar bound
in \Cref{thm:main}. For fixed $p$, the sampling MSE is
$\mathcal O_T(N^{-4/3})$.
{{The exponent is independent of $p$, but the sum of constants in
\eqref{eq:product-mwg-rate} need not be bounded uniformly in $p$.}}
\end{corollary}
\begin{proof}
See \Cref{subsec:supp-product-mwg-proof} of the supplementary materials.
\end{proof}

\begin{corollary}[Additive polynomial tests]
\label{cor:polynomial-tests}
For the product/MWG construction above, suppose $\mathcal X_r=\mathbb R$,
all particles start at the same deterministic $\bm{x}_0$ with
$\pi_r(x_{0,r})>0$, and each coordinatewise one-dimensional MH construction satisfies the sampling and kernel assumptions of \Cref{thm:main}.
Use $Y_{r,j}(x,u)=x+\sigma_r\Phi^{-1}(u)$, where $0<u<1$,
$\sigma_r>0$ is fixed across sweeps and particle counts, and $\Phi$ is
the standard normal CDF. For fixed integers $0\le T_0<T$ and
$\ell_r\ge1$, let $f(\bm{x})=\sum_{r=1}^p c_r x_r^{\ell_r}$ with
deterministic $c_r$. Then, for all sufficiently large admissible $N$,
\begin{equation}\label{eq:additive-polynomial-rate}
 \mathbb E_{\mathrm{RQMC}}
 [(\widehat\mu_{N,T}(f)-\mu_T(f))^2]
 \le N^{-4/3}\sum_{r=1}^p c_r^2{{K_{r,T_0,T,\ell_r}}}(\log N)^{\ell_r},
\end{equation}
where ${{K_{r,T_0,T,\ell_r}}}<\infty$ depends on the corresponding coordinate target and proposal, initial
state, $T,T_0,\ell_r,b,t$, but not on $N$.
For fixed $p$ and $\ell_r=\ell$ for all $r$, the bound is
$\mathcal O_{T,\ell}(N^{-4/3}(\log N)^\ell)$.
For $p=1$, this recovers the scalar polynomial extension {{of \Cref{prop:scalar-polynomial} in the supplementary materials}}.
\end{corollary}
\begin{proof}
See \Cref{sec:supp-polynomial} of the supplementary materials.
\end{proof}

\Needspace{6\baselineskip}
\section{Numerical experiments}
\label{sec:numerics}

We compare MH driven by IID inputs with the Array-RQMC estimator $\widehat\mu_{N,T}(f)$, targeting
\(\mu_T(f)\), the average over the finite horizon defined in \Cref{sec:main}. 
The one-dimensional examples in \Cref{sec:verified-1d} satisfy the
conditions of \Cref{thm:main}. The product Gaussian coordinatewise-MWG
construction in \Cref{sec:gaussian-multidim} is covered by
\Cref{cor:polynomial-tests}.
{{For the joint Hilbert-MH experiments on the Gaussian target
and the Bayesian logistic posteriors in \Cref{sec:bayesian},
the projection tests are unbounded, so \Cref{thm:md-fixed}
does not apply as stated. Moreover, neither the uniform tube
condition nor the propagated-HK condition has been verified
for these examples.}}
Thus the slopes
$-1-1/(p+2)$ shown for these joint updates are {{nominal exponents}},
not proved rates for the reported tests.

\subsection{Experimental protocol}
We use MATLAB's \texttt{MatousekAffineOwen} scramble for $N=2^m$
Sobol' points {{with
\texttt{Skip=0}, \texttt{Leap=0}, and
\texttt{PointOrder=graycode}}}, applying a fresh independent randomization at each transition \citep{hong2003algorithm,owen2003variance}.
For a fixed net, this randomization has the same quadrature variance as
nested uniform scrambling {{\citep[Section~2.3]{owen2008local}}}.
{{For joint Hilbert-MH, we use Skilling's axes-to-transpose
construction \citep{skilling2004hilbert}, with
$B_p=\min\{20,\lfloor64/p\rfloor\}$ bits per coordinate.}}
Empirical variances are sample variances across $R_{\rm rep}=200$
independent macro-replications with $T_0=8$ and $T=128$ for the
reported experiments. These horizons count scalar MH transitions in
one dimension and joint MH transitions in the Hilbert experiments;
for coordinatewise MWG, horizons count complete sweeps and each
coordinate update uses a separate independent scramble.

We fit log--log variance slopes by
ordinary least squares over the largest five particle counts.
Writing $\widehat V_N^{\rm IID}$ and $\widehat V_N^{\rm Array}$ for the
empirical variances, the variance-reduction factor (VRF) is
$\widehat{\mathrm{VRF}}_N=\widehat V_N^{\rm IID}/\widehat V_N^{\rm Array}$.
{{We report 95\% percentile bootstrap confidence intervals
for the fitted slopes and VRFs at the largest particle
count, based on 1,000 resamples of the independent
macro-replications. Resampling is performed independently
across particle counts and methods, while estimates
computed from the same particle trajectories retain
their common resampling indices.
To assess computational efficiency, we separately time
complete serial runs of each method, using 20 repetitions
and the median elapsed time.
The time-adjusted efficiency (TAE) is defined as
\(
\widehat{\mathrm{TAE}}_N
=
\widehat{\mathrm{VRF}}_N\times
\widetilde t_N^{\rm IID}/
{\widetilde t_N^{\rm Array}},
\)
where $\widetilde t_N^{\rm IID}$ and
$\widetilde t_N^{\rm Array}$ denote the respective
median runtimes. We keep $T_0$ and $T$ fixed to study how sampling
variance depends on $N$. Both methods target the same
finite-horizon expectation $\mu_T(f)$ and therefore have
the same bias relative to $\Pi(f)$, independent of $N$.
The reported VRFs and TAEs assess sampling-variance
reduction and its computational cost, rather than MSE
relative to the stationary target.}}

\subsection{Verified one-dimensional examples}
\label{sec:verified-1d}
We use $N=2^{10},\ldots,2^{18}$, $x_0=0$ and the
Gaussian random-walk proposal $Y=x+2.38\Phi^{-1}(u)$.
\Cref{tab:numerical-targets} lists the four targets.
For each target $\pi$, the CDF test is
$f_{\mathrm{CDF}}(x)=\one\{x\le a_\pi\}$, where
$a_\pi=F_\pi^{-1}(1/4)$ is obtained by adaptive quadrature and root
finding. This test has variation one and directly tests the
$\mathcal O_T(N^{-4/3})$ bound in \Cref{thm:main}, whose conditions are
verified in \Cref{sec:verification}. We also use $f(x)=x$ and $f(x)=x^2$;
taking $p=1$ in \Cref{cor:polynomial-tests} covers these tests with the respective
fixed-horizon bounds $\mathcal O_T(N^{-4/3}\log N)$ and
$\mathcal O_T(N^{-4/3}(\log N)^2)$.
{{For the Student-$t_3$ target, the stationary fourth moment is
infinite, so $X^2$ has infinite stationary variance. At any fixed
horizon from the deterministic start, Gaussian proposal increments
give finite moments of every order; the finite-horizon polynomial
bound does not assert finite stationary variance.}}

\begin{table}[htbp]
\centering
\caption{Target distributions used in the one-dimensional experiments;
log densities are given up to an additive constant.}
\label{tab:numerical-targets}
\begin{tabular}{@{}lll@{}}
\toprule
Target & Log density $\log\pi(x)$ & Main feature\\
\midrule
Standard normal (SN)  & $-x^2/2$ & log-concave benchmark\\
Double well (DW)  & $-(x^2-1)^2$ & symmetric bimodality\\
Gaussian mixture (GM) & $\log[e^{-2(x+1.5)^2}+0.8e^{-3x^2}+0.6e^{-4(x-1.5)^2}]$ & asymmetric multimodality\\
Student-$t_3$ (ST) & $-2\log(1+x^2/3)$ & heavy tails\\
\bottomrule
\end{tabular}
\end{table}

\begin{table}[t]
\centering
\caption{{{VRFs, fitted Array-RQMC
variance slopes, and TAEs for the
one-dimensional experiments. VRFs and TAEs are evaluated at
$N=2^{18}$; slopes are fitted over $N=2^{14},\ldots,2^{18}$. Bracketed values below estimates are 95\% percentile bootstrap
confidence intervals.}}}
\label{tab:all-vrf-slopes}

\begingroup
\small
\setlength{\tabcolsep}{5pt}
\renewcommand{\arraystretch}{1}

\newcommand{\estimateci}[2]{\begingroup
  \renewcommand{\arraystretch}{0.8}\begin{tabular}[t]{@{}c@{}}
    $#1$\\[-2pt]
    {\scriptsize $#2$}
  \end{tabular}\endgroup
}

\scalebox{0.95}{{{\begin{tabular}{@{}l*{9}{c}@{}}
\toprule
& \multicolumn{3}{c}{$f_{\mathrm{CDF}}$}
& \multicolumn{3}{c}{$x$}
& \multicolumn{3}{c}{$x^2$}\\
\cmidrule(lr){2-4}
\cmidrule(lr){5-7}
\cmidrule(lr){8-10}
Target
& VRF & Slope & TAE
& VRF & Slope & TAE
& VRF & Slope & TAE\\
\midrule

SN
& \estimateci{50.31}{[37.58,68.70]}
& \estimateci{-1.35}{[-1.43,-1.26]}
& $27.12$
& \estimateci{33.48}{[25.32,44.80]}
& \estimateci{-1.33}{[-1.42,-1.24]}
& $18.38$
& \estimateci{26.31}{[20.40,34.52]}
& \estimateci{-1.39}{[-1.48,-1.29]}
& $14.14$\\
\addlinespace[0.6em]

DW
& \estimateci{51.44}{[37.86,67.76]}
& \estimateci{-1.35}{[-1.44,-1.25]}
& $26.73$
& \estimateci{48.02}{[36.29,64.32]}
& \estimateci{-1.32}{[-1.42,-1.22]}
& $25.57$
& \estimateci{33.52}{[26.23,42.82]}
& \estimateci{-1.30}{[-1.38,-1.21]}
& $17.46$\\
\addlinespace[0.6em]

GM
& \estimateci{37.68}{[28.76,49.45]}
& \estimateci{-1.33}{[-1.42,-1.25]}
& $23.45$
& \estimateci{36.03}{[28.29,45.80]}
& \estimateci{-1.29}{[-1.37,-1.21]}
& $22.81$
& \estimateci{36.73}{[27.45,48.55]}
& \estimateci{-1.41}{[-1.50,-1.31]}
& $22.83$\\
\addlinespace[0.6em]

ST
& \estimateci{51.47}{[38.85,67.64]}
& \estimateci{-1.33}{[-1.42,-1.25]}
& $28.78$
& \estimateci{30.43}{[22.95,39.71]}
& \estimateci{-1.36}{[-1.45,-1.28]}
& $17.29$
& \estimateci{18.08}{[13.69,24.17]}
& \estimateci{-1.53}{[-1.63,-1.42]}
& $10.06$\\

\bottomrule
\end{tabular}}}}
\endgroup
\end{table}

{{\Cref{tab:all-vrf-slopes} shows gains for all twelve
target--test combinations at $N=2^{18}$, with VRFs of
$18.1$--$51.5$ and TAEs of $10.1$--$28.8$.
The CDF slopes range from $-1.35$ to $-1.33$, and all four
95\% confidence intervals contain $-4/3$, consistent with
the verified bound. For the polynomial tests, the fitted slopes range from
$-1.53$ to $-1.29$. In particular, the $x^2$ test under the Student-$t_3$ target
has a fitted slope of $-1.53$, with a 95\% confidence interval
of $[-1.63,-1.42]$. This apparent decay faster than
$N^{-4/3}(\log N)^2$ is observed only over the finite range
of particle counts considered here; the theoretical result
also provides an upper bound rather than an exact asymptotic rate.}}

\subsection{Product Gaussian benchmark: coordinatewise MWG and joint Hilbert MH}
\label{sec:gaussian-multidim}
We consider the product target $\mathcal N_p(\bm{0},I_p)$ for $p=2,4,8$
and $f_p(\bm{x})=p^{-1/2}\sum_{r=1}^p x_r$.
The coordinatewise construction uses systematic-scan MWG with
$Y_r=x_r+2.38\Phi^{-1}(u)$, sorting by the updated coordinate and using
a fresh independently scrambled three-dimensional Sobol' point set at
every coordinate update. The joint construction uses
$\bm{Y}=\bm{x}+(2.38/\sqrt p)\bm{Z}$, with
$\bm{Z}\sim\mathcal N_p(\bm{0},I_p)$, Hilbert sorting and point-set
dimension $D=p+2$. Its sorting map is
$\bm{\psi}(\bm{x})=(\Phi(x_1),\ldots,\Phi(x_p))$, {{with Hilbert-index precision as specified in the experimental protocol}}.
Both methods start every particle at $\bm{0}$, use
$N=2^{10},\ldots,2^{18}$ and the replication counts and horizons in the
experimental protocol, and are compared with their matching IID
kernels. MWG records complete sweep endpoints as in \Cref{sec:product-mwg}.

\begin{figure}[htbp]\centering
\includegraphics[width=\textwidth]{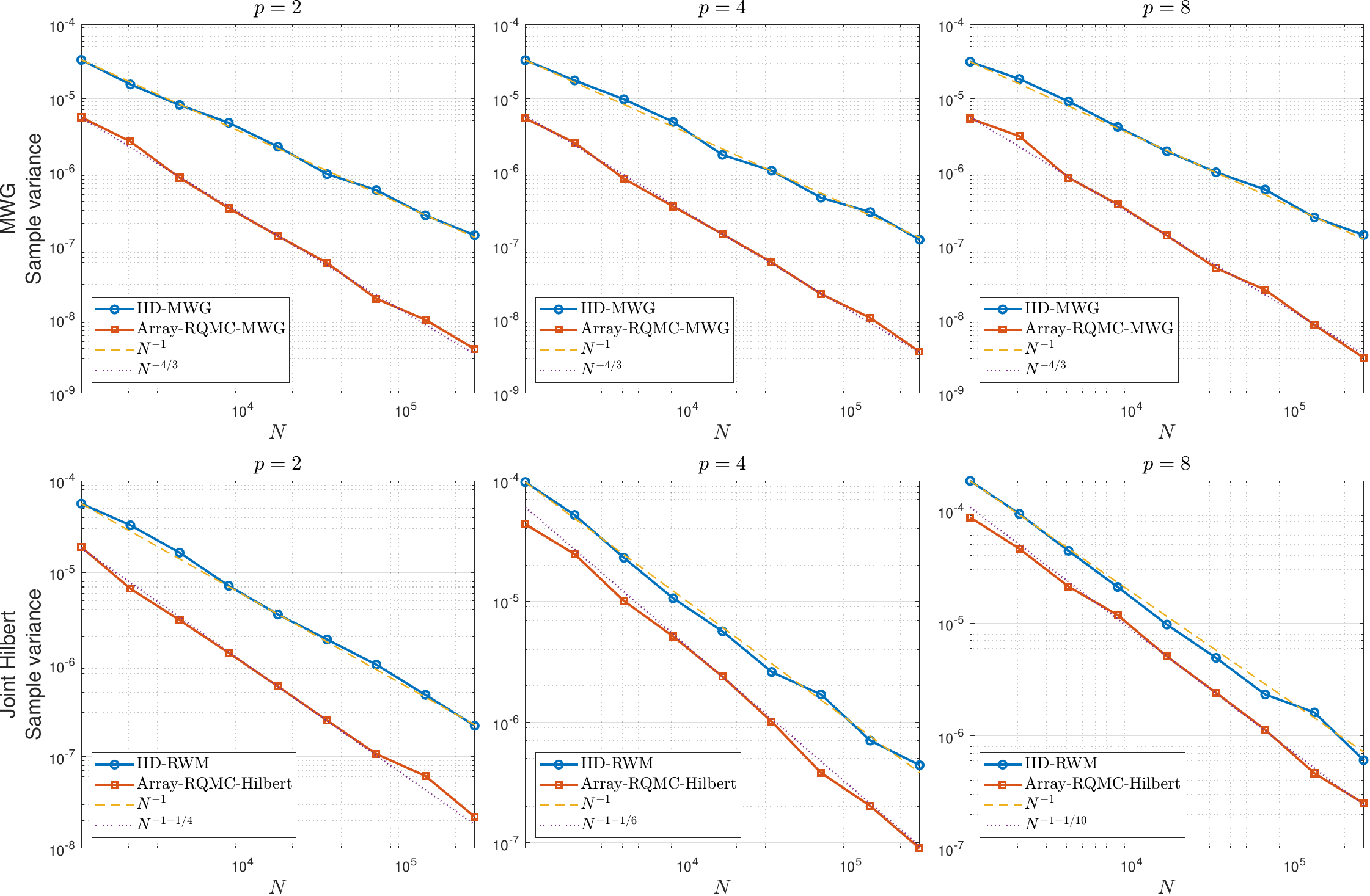}
\caption{Standard Gaussian time-average sampling variances for
$f_p(\bm{x})=p^{-1/2}\sum_{r=1}^p x_r$ in $p=2,4,8$ dimensions
(left to right). Top: coordinatewise MWG; bottom: joint Hilbert MH.
Dashed lines have slope $-1$; dotted lines show the $-4/3$ power
reference for MWG and {{the nominal exponent}} $-1-1/(p+2)$ for joint Hilbert MH. {{In}} {{the legends, IID-MWG and IID-RWM denote the matching IID kernels, and Array-RQMC-MWG and Array-RQMC-Hilbert denote coordinatewise MWG and joint Hilbert MH; the vertical axes show the replicate sample variance of the time average.}}}
\label{fig:gaussian-multidim}
\end{figure}

\begin{table}[htbp]
\centering
\caption{{{VRFs, fitted Array-RQMC
variance slopes, and TAEs for the
standard Gaussian benchmark.
VRFs and TAEs are evaluated at $N=2^{18}$; slopes are fitted
over $N=2^{14},\ldots,2^{18}$.
Each method is compared with its matching IID kernel.}}}
\label{tab:gaussian-mwg-hilbert}

\begingroup
\small
\setlength{\tabcolsep}{4pt}
\renewcommand{\arraystretch}{1}

\newcommand{\estimateci}[2]{\begingroup
  \renewcommand{\arraystretch}{0.8}\begin{tabular}[t]{@{}c@{}}
    $#1$\\[-2pt]
    {\scriptsize $#2$}
  \end{tabular}\endgroup
}
\newcommand{\headerci}[1]{\begingroup
  \renewcommand{\arraystretch}{0.8}\begin{tabular}[t]{@{}c@{}}
    #1\\[-2pt]
    {\scriptsize (95\% CI)}
  \end{tabular}\endgroup
}

\scalebox{0.93}{{{\begin{tabular}{@{}l*{9}{c}@{}}
\toprule
& \multicolumn{3}{c}{$p=2$}
& \multicolumn{3}{c}{$p=4$}
& \multicolumn{3}{c}{$p=8$}\\
\cmidrule(lr){2-4}
\cmidrule(lr){5-7}
\cmidrule(lr){8-10}
Method
& \headerci{VRF} & \headerci{Slope} & TAE
& \headerci{VRF} & \headerci{Slope} & TAE
& \headerci{VRF} & \headerci{Slope} & TAE\\
\midrule

MWG
& \estimateci{35.02}{[25.75,46.62]}
& \estimateci{-1.28}{[-1.37,-1.18]}
& $14.36$
& \estimateci{33.18}{[24.87,43.79]}
& \estimateci{-1.31}{[-1.40,-1.22]}
& $12.72$
& \estimateci{46.39}{[35.57,60.35]}
& \estimateci{-1.36}{[-1.44,-1.27]}
& $15.72$\\
\addlinespace[0.6em]

Hilbert MH
& \estimateci{9.82}{[7.45,12.79]}
& \estimateci{-1.14}{[-1.24,-1.05]}
& $3.45$
& \estimateci{4.89}{[3.68,6.62]}
& \estimateci{-1.18}{[-1.27,-1.09]}
& $1.79$
& \estimateci{2.42}{[1.78,3.22]}
& \estimateci{-1.11}{[-1.19,-1.01]}
& $0.99$\\

\bottomrule
\end{tabular}}}}
\endgroup
\end{table}

\Cref{tab:gaussian-mwg-hilbert,fig:gaussian-multidim} show that the
coordinatewise-MWG slopes remain near the $-4/3$ power reference, {{with all three 95\% confidence
intervals containing $-4/3$.  Neither the VRFs nor the TAEs show a clear deterioration across $p=2,4,8$, and the TAEs range from $12.72$ to $15.72$.}} Exploiting product structure keeps each RQMC update
three-dimensional. For the linear test $f_p$,
\Cref{cor:polynomial-tests} with $\ell_r=1$ and
$c_r=p^{-1/2}$ gives the
$\mathcal O_T(N^{-4/3}\log N)$ bound, whose power exponent
is independent of dimension.

{{For joint Hilbert MH, the VRFs decrease substantially
with dimension, while the TAEs fall from $3.45$ at $p=2$
to $0.99$ at $p=8$.
Thus, at $p=8$, the variance reduction is approximately
offset by the additional computational cost. The lines with slopes $-1-1/(p+2)$ serve only as nominal
references: the theorem's hypotheses remain unverified
for these joint-update examples, and agreement with a
reference slope does not establish them.}}

\subsection{\texorpdfstring{{{Bayesian logistic regression}}}{Bayesian logistic regression}}
\label{sec:bayesian}
\begin{figure}[htbp]\centering
\includegraphics[width=0.8\textwidth]{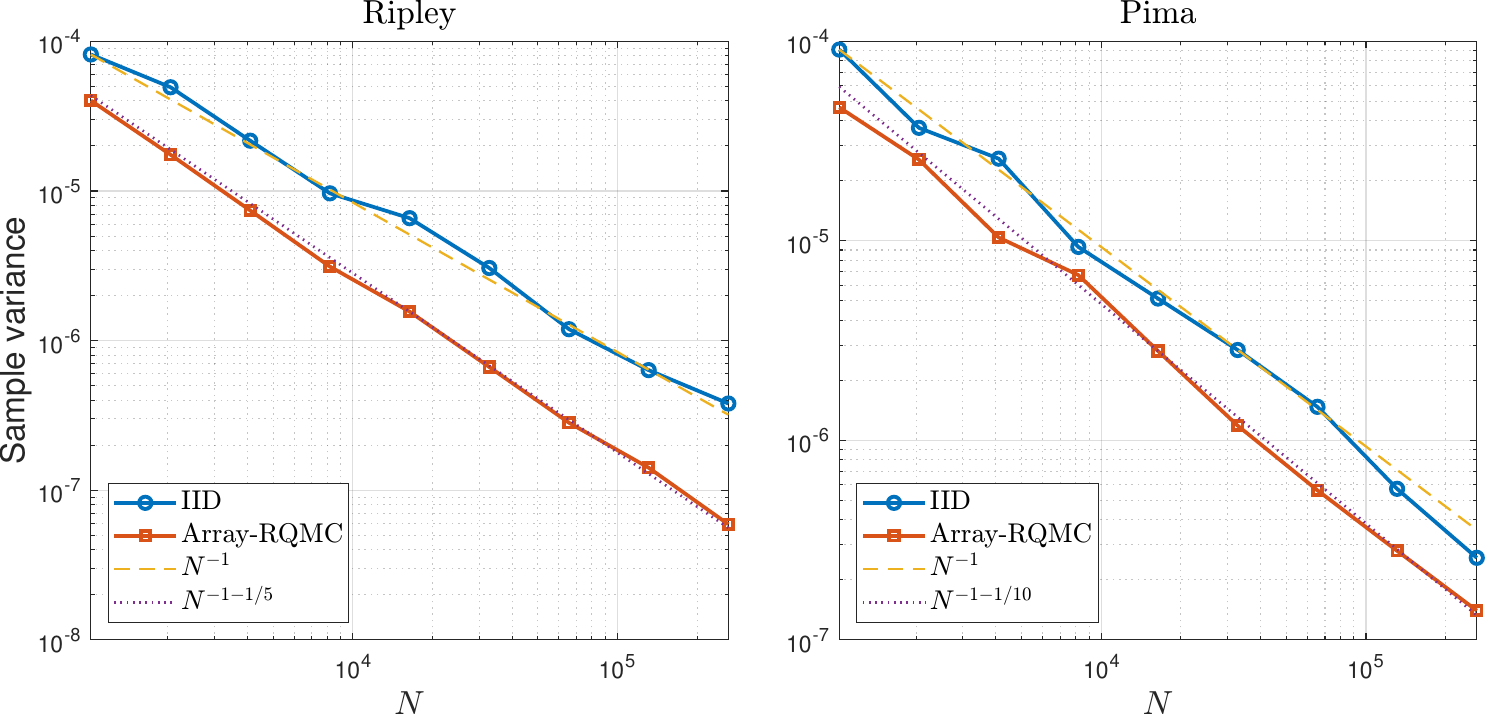}
\caption{Time-average sampling variance for $f_{\mathrm{proj}}(\bm{\beta})$ in Ripley
(left, $p=3$) and Pima (right, $p=8$). Dashed lines have slope $-1$;
dotted lines show the {{nominal}} slopes $-1.2$ and $-1.1$,
respectively.}
\label{fig:ripley-variance}
\end{figure}

These dependent posterior examples use joint Hilbert-sorted MH. We use the Ripley ($n=250$, $p=3$) and Pima
($n=532$, $p=8$) logistic-regression models of \citet{alexopoulos2023variance}, where $n$ denotes the
number of observations and $p$ the number of regression
coefficients, including the intercept. For binary response $y_r$ and design vector
$\bm{a}_r\in\mathbb R^p$, the log posterior under a flat prior is, up to
an additive constant,
\[
 \ell(\bm{\beta})=\sum_{r=1}^{n}
 \{y_r\bm{a}_r^\top\bm{\beta}-\log(1+\exp(\bm{a}_r^\top\bm{\beta}))\}.
\]
Both methods start every particle at the deterministic MLE
$\widehat{\bm{\beta}}$.
Writing $\widehat\Sigma=[-\nabla^2\ell(\widehat{\bm{\beta}})]^{-1}$ and
$\widehat L\widehat L^\top=\widehat\Sigma$, the proposal is
$\bm{\beta}'=\bm{\beta}+(2.38/\sqrt p)\widehat L\bm{Z}$ with
$\bm{Z}\sim\mathcal N_p(\bm{0},I_p)$.
Hilbert sorting uses the marginal-standardized map
$\psi_j(\bm{\beta})=\Phi((\beta_j-\widehat\beta_j)/
\sqrt{\widehat\Sigma_{jj}})$; the Sobol' dimensions are $D=5$ and $D=10$.
We use $N=2^{10},\ldots,2^{18}${{, with $T_0=8$, $T=128$ and $R_{\rm rep}=200$ as in the experimental protocol}}.
{{\Cref{fig:ripley-variance} shows time-average sampling
variances for the standardized linear projection
\(
f_{\mathrm{proj}}(\bm{\beta})
=
\frac{1}{\sqrt p}
\sum_{j=1}^{p}
\frac{\beta_j-\widehat\beta_j}
{\sqrt{\widehat\Sigma_{jj}}}.
\)
This is the analogue of the Gaussian projection test, with
centering at the MLE and scaling by the diagonal of the inverse
observed information.}}

\begin{table}[htbp]
\centering
\caption{{{VRFs, fitted
Array-RQMC variance slopes, and
TAEs for the Bayesian
logistic-regression experiments.
VRFs and TAEs are evaluated at $N=2^{18}$; slopes are fitted
over $N=2^{14},\ldots,2^{18}$.}}}
\label{tab:bayesian-projection}

\begingroup
\small
\setlength{\tabcolsep}{9pt}
\renewcommand{\arraystretch}{1}

\scalebox{1}{{{\begin{tabular}{@{}lcccc@{}}
\toprule
Dataset & $p$
& VRF {\scriptsize (95\% CI)}
& Slope {\scriptsize (95\% CI)}
& TAE\\
\midrule
Ripley & $3$
& $6.42$ {\scriptsize $[4.94,8.28]$}
& $-1.17$ {\scriptsize $[-1.26,-1.07]$}
& $5.97$\\
Pima & $8$
& $1.84$ {\scriptsize $[1.37,2.44]$}
& $-1.07$ {\scriptsize $[-1.17,-0.97]$}
& $1.72$\\
\bottomrule
\end{tabular}}}}

\endgroup
\end{table}

{{\Cref{tab:bayesian-projection,fig:ripley-variance}
show fitted Array-RQMC slopes of $-1.17$ for Ripley and
$-1.07$ for Pima. As in the joint Hilbert-MH Gaussian benchmark, the lower-dimensional example shows a clearer improvement over the Monte Carlo slope $-1$, whereas the gain is weaker at $p=8$. Moreover, unlike the Gaussian experiments, the Bayesian TAEs
remain close to their VRFs: $5.97$ versus $6.42$ for Ripley
and $1.72$ versus $1.84$ for Pima.
Hilbert MH takes only about $7.5\%$ and $6.6\%$ longer
than IID MH, respectively, suggesting that the additional
cost of sorting and related operations is small relative
to the total runtime.}}

\section{Discussion and conclusions}
\label{sec:conclusion}
{{The verified one-dimensional result gives the $4/3$ sampling-MSE
exponent for pointwise BV tests. The same exponent holds for additive
marginal BV tests on product targets when coordinatewise MWG reduces
each update to a three-dimensional Array-RQMC problem; its constants
may grow with dimension. For Gaussian proposals and additive
polynomials of maximum degree $\ell$, the bound includes
$(\log N)^\ell$.}}
The Gaussian experiments reflect this distinction: coordinatewise
MWG retains slopes near $-4/3$ across $p=2,4,8$, whereas the gain
from joint Hilbert-sorted MH weakens with dimension in these experiments.

{{For joint $p$-dimensional MH, the exponent $1+1/(p+2)$ belongs
to a conditional framework. Uniform boundary-tube bounds and the
propagated-HK condition are not consequences of Hilbert ordering and
remain unverified for the joint examples.
Obtaining checkable assumptions for dependent
multidimensional MH remains open.}} Existing composition criteria for
HK variation, such as \citet[Theorem~4]{basu2016}, may help with the
propagated-test condition, but do not directly cover the unbounded
Gaussian-CDF setting considered here. {{The plotted joint-update
exponents are therefore heuristic comparisons, not established
convergence rates for those experiments.}}

The theory keeps $T$ fixed and does not provide a joint $(N,T)$ rate
for the whole-line random-walk examples. The global BV-contraction route
used in \Cref{rem:horizon-growth} is ruled out in that setting, whereas
the supplementary horizon experiment shows empirical variance decay
close to $L^{-1}$ and is consistent with an
$N^{-4/3}L^{-1}$ scaling. Bridging this gap is also a future research direction.

The product reduction relies on coordinate proposals and acceptance
probabilities that depend only on the updated coordinate. For dependent
targets, other coordinates generally enter the acceptance decision,
and the marginal-error independence used in \Cref{cor:product-mwg}
need not hold. Verifiable theory for such dependent updates, together
with sharper control of dimension and effective dimension, remains an
open direction alongside joint $(N,T)$ bounds.

\appendix
\crefalias{section}{appendix}

\section{Proof of \texorpdfstring{\Cref{thm:md-fixed}}{Theorem \getrefnumber{thm:md-fixed}}}
\label{app:md-proof}
We first prove the common scrambled-net tube bound and exact local
orthant representation. The conditional orthant-to-HK transfer then
controls test-function errors. Dimension-independent innovation
identities and orthogonality complete the proof of \Cref{thm:md-fixed}.

\subsection{Scrambled-net tube-to-variance bound}
\label{app:variance}
The indicator exponent in \citet[Theorem 4.4]{he2015convergence}
motivates the following dimension-dependent bound. We use dimension $D=p+2$
for the MH framework of this paper, dimension three for one-dimensional MH and dimension two
for one-dimensional inversion. The proof uses only exact net counts and a standard scrambled-net gain
bound, and requires no Minkowski-content limit.
For $A\subset\mathbb R^D$, write
$A_\varepsilon=\{\bm{y}:\operatorname{dist}_2(\bm{y},A)\le\varepsilon\}$,
with $\operatorname{dist}_2(\bm{y},\varnothing)=\infty$.
Boundaries and tubes are taken in the ambient Euclidean space.

For a nested uniform scramble of a $(t,m,D)$-net, $m\ge t$, the gain bound
{{\citet[eq.~(4) and Section~2.4]{PanOwen2023}}} gives, for every
$g\in L^2([0,1)^D)$,
\begin{equation}\label{eq:scrambled-gain}
 \operatorname{Var}(Q_Ng)
 \le\frac{G_{b,t,D}}N\left\|g-\int_{[0,1)^D}g\right\|_2^2,
 \qquad G_{b,t,D}=b^t\left(\frac{b+1}{b-1}\right)^D.
\end{equation}
This conservative choice follows from the gain estimates of Owen summarized
there; it is uniform in $m$, the underlying net and $g$.

\begin{proposition}[Uniform tube-to-variance bound]
\label{thm:tube-variance}
Fix $b\ge2$, $t\ge0$ and $D\ge2$. For $N=b^m$, $m\ge t$, let $Q_N$ be the equal-weight
quadrature rule based on a nested uniform scramble of a
$(t,m,D)$-net in base $b$. Let $B\subset[0,1)^D$ be measurable
and deterministic relative to the scramble. Suppose
\[
 \lambda_D((\partial B)_\varepsilon)\le M\varepsilon,
 \qquad 0<\varepsilon<\varepsilon_0,
\]
where $M\ge0$ and $\varepsilon_0>0$. Then
\begin{equation}\label{eq:tube-variance-constant}
 \operatorname{Var}(Q_N\one_B)\le c_{b,t,D}M N^{-1-1/D},
 \qquad c_{b,t,D}:=\tfrac14G_{b,t,D}\sqrt{D}\,b^{1+t/D},
\end{equation}
for every admissible $N=b^m\ge N_0$, where one may take
\begin{equation}\label{eq:tube-cutoff}
 k_0=\max\{0,\lfloor\log_b(\sqrt{D}/\varepsilon_0)\rfloor+1\},
 \qquad N_0=b^{t+D k_0}.
\end{equation}
\end{proposition}
\begin{proof}
Put $k=\lfloor(m-t)/D\rfloor$ and $h=b^{-k}$. Partition $[0,1)^D$ into
half-open base-$b$ cubes $\mathcal C$ of side $h$. Since $Dk\le m-t$, each cube
is a union of $b^{m-t-Dk}$ elementary intervals of volume $b^{t-m}$.
It therefore contains exactly $N|\mathcal C|$ net points almost surely. Set
$p_{\mathcal C}=|\mathcal C|^{-1}\int_{\mathcal C}\one_B$ and let $f_k=p_{\mathcal C}$ on $\mathcal C$. Exact integration
of these cell averages gives $Q_Nf_k=\int f_k=\lambda_D(B)$.
For $r_B=\one_B-f_k$ we consequently have
\[
 \int r_B=0,\qquad
 \operatorname{Var}(Q_N\one_B)=\operatorname{Var}(Q_Nr_B),\qquad
 \|r_B\|_2^2=\sum_{\mathcal C}|\mathcal C|p_{\mathcal C}(1-p_{\mathcal C}).
\]
Only cells with $0<p_{\mathcal C}<1$ contribute. Each such cell meets both $B$ and
its complement, so its connectedness implies that it meets $\partial B$.
Its diameter is at most $\sqrt{D}h$, and hence the union of contributing
cells lies in $(\partial B)_{\sqrt{D}h}$. Thus, if $\sqrt{D}h<\varepsilon_0$,
\[
 \|r_B\|_2^2\le\tfrac14\lambda_D((\partial B)_{\sqrt{D}h})
 \le\tfrac14M\sqrt{D}h.
\]
Null exceptional subsets in constant cells have no $L^2$ contribution
and are missed almost surely by the finitely many marginally uniform points.
Applying \eqref{eq:scrambled-gain} to $r_B$ and using
$h\le b^{1+t/D}N^{-1/D}$ gives
$$
\operatorname{Var}(Q_N\mathbf1_B)
\le
\frac{G_{b,t,D}}{N}\|r_B\|_{2}^2
\le
\frac14G_{b,t,D}M\sqrt{D}\,\frac{h}{N}
\le
c_{b,t,D}M N^{-1-1/D}.
$$
The cutoff \eqref{eq:tube-cutoff} ensures
$k\ge k_0$ and $\sqrt{D}h<\varepsilon_0$, including exact powers in the
logarithm. Empty sets and $M=0$ satisfy the same weak inequalities.
\end{proof}

For \Cref{thm:md-fixed}, one may take
$C_{b,t,D}=c_{b,t,D}=\tfrac14G_{b,t,D}\sqrt D\,b^{1+t/D}$.
Applying \eqref{eq:tube-cutoff} with radius $\varepsilon_{0,s}$ gives
\begin{equation}\label{eq:md-cutoff}
 k_{0,s}:=\max\{0,\lfloor\log_b(\sqrt D/\varepsilon_{0,s})\rfloor+1\},
 \qquad N_{0,s}:=b^{t+Dk_{0,s}}.
\end{equation}
The bound holds for every admissible $N\ge\max_{s\le T}N_{0,s}$.

\subsection{Exact local orthant representation}
\label{app:md-local}
For a deterministic Borel $A\subseteq\mathcal X$, define, only for use
in the proof,
\begin{equation}\label{eq:md-generic-event}
 \begin{aligned}
 B_{N,j,A}&:=\{(v,\bm{u},w)\in[0,1)^D:
                \bm{\Psi}_j(\bm{Z}_{N,j-1}(v),\bm{u},w)\in A\},\\
 {{e_{N,j}}}(A)&:=\widehat\nu_{N,j}(A)
                   -\widehat\nu_{N,j-1}(\mathsf P_j\one_A).
 \end{aligned}
\end{equation}
For $A=A_{\bm{x}}$ this gives $B_{N,j,\bm{x}}$ from \eqref{eq:md-local-event}.
Write $Q_{N,j}g=N^{-1}\sum_{k=0}^{N-1}g(\bm{W}_{k,j})$, using the
complete points defined in \Cref{ass:points} and their original labels.

\begin{lemma}[Exact local event representation and orthant bound]
\label{lem:md-local}
For the Array-RQMC update \eqref{eq:array-update} with
point sets satisfying \Cref{ass:points}, fix $1\le j\le T$. 
For every deterministic Borel set $A\subseteq\mathcal X$,
almost surely,
\begin{equation}\label{eq:md-local-identity}
\begin{aligned}
 {{e_{N,j}}}(A)
 =Q_{N,j}\one_{B_{N,j,A}}-\lambda_D(B_{N,j,A}),\qquad
 \mathbb E_{\mathrm{RQMC}}
 [{{e_{N,j}}}(A)\mid\mathcal F_{N,j-1}]
 =0.
\end{aligned}
\end{equation}

For the lower orthants
$A_{\bm{x}}=\mathcal X\cap
\prod_{r=1}^p(-\infty,x_r]$,
suppose that deterministic constants $0\le M_j<\infty$
and $\varepsilon_{0,j}>0$ satisfy
\begin{equation}\label{eq:md-local-tube}
 \lambda_D((\partial B_{N,j,\bm{x}})_\varepsilon)
 \le M_j\varepsilon,
 \qquad 0<\varepsilon<\varepsilon_{0,j},
\end{equation}
uniformly over $\bm{x}\in\mathbb R^p$, admissible $N$,
and realized old particle arrays.
Then, for every $\bm{x}\in\mathbb R^p$ and every
admissible $N\ge N_{0,j}$, with $N_{0,j}$ given by
\eqref{eq:md-cutoff} at $s=j$,
\begin{equation}\label{eq:md-local-variance}
 \mathbb E_{\mathrm{RQMC}}
 [{{e_{N,j}}}(A_{\bm{x}})^2
  \mid\mathcal F_{N,j-1}]
 \le c_{b,t,D}M_jN^{-1-1/D}.
\end{equation}
The constant and cutoff are uniform in $\bm{x}$, $N$,
and the old particle array. The centering identity in
\eqref{eq:md-local-identity} holds for every admissible $N$
without the tube condition.
\end{lemma}

\begin{proof}
Fix a history through time $j-1$. The old particle array,
its sorting permutation, and $\bm{Z}_{N,j-1}$ are then fixed.
The fresh scramble is independent of this history. First, let $A\subseteq\mathcal X$ be any deterministic
Borel set. By exact first-coordinate stratification,
$V_{(i),j}\in[i/N,(i+1)/N)$ for every $i$, almost surely.
Consequently,
\[
\begin{aligned}
 Q_{N,j}\one_{B_{N,j,A}}
 &=\frac1N\sum_{i=0}^{N-1}
   \one\bigl\{
   \bm{\Psi}_j(
      \bm{Z}_{N,j-1}(V_{(i),j}),
      \bm{U}_{(i),j},\Omega_{(i),j})
   \in A\bigr\}\\
 &=\frac1N\sum_{i=0}^{N-1}
   \one\bigl\{
   \bm{\Psi}_j(
      \bm{X}_{\varsigma_j(i),j-1},
      \bm{U}_{(i),j},\Omega_{(i),j})
   \in A\bigr\}=\widehat\nu_{N,j}(A).
\end{aligned}
\]

On the other hand, integrating over the transition
inputs gives
\[
\begin{aligned}
 \lambda_D(B_{N,j,A})
 &=\int_0^1\int_{[0,1)^p}\int_0^1
   \one_A\bigl(
     \bm{\Psi}_j(
       \bm{Z}_{N,j-1}(v),\bm{u},w)
   \bigr)
   \,\dd w\,\dd\bm{u}\,\dd v\\
 &=\int_0^1
   \mathsf P_j(\bm{Z}_{N,j-1}(v),A)\,\dd v=\frac1N\sum_{i=0}^{N-1}
   \mathsf P_j(
     \bm{X}_{\varsigma_j(i),j-1},A)=\widehat\nu_{N,j-1}(\mathsf P_j\one_A).
\end{aligned}
\]
Subtracting establishes the first identity in
\eqref{eq:md-local-identity}. Conditional on $\mathcal F_{N,j-1}$, the event region
$B_{N,j,A}$ is deterministic. Each original, unsorted
point $\bm{W}_{k,j}$ is marginally uniform on
$[0,1)^D$. Hence
\[
\begin{aligned}
 &\mathbb E_{\mathrm{RQMC}}
 [Q_{N,j}\one_{B_{N,j,A}}
  \mid\mathcal F_{N,j-1}]=
 \frac1N\sum_{k=0}^{N-1}
 \mathbb E_{\mathrm{RQMC}}
 [\one_{B_{N,j,A}}(\bm{W}_{k,j})
  \mid\mathcal F_{N,j-1}]=\lambda_D(B_{N,j,A}).
\end{aligned}
\]
This proves conditional centering without requiring
the individually sorted points to be uniform. Now take $A=A_{\bm{x}}$, so that
$B_{N,j,A}=B_{N,j,\bm{x}}$.
Under \eqref{eq:md-local-tube}, the event region
satisfies the assumptions of \Cref{thm:tube-variance},
conditional on the history. Since the fresh scramble
has its original distribution conditionally on
$\mathcal F_{N,j-1}$, that proposition yields,
for every admissible $N\ge N_{0,j}$,
\[
\begin{aligned}
 &\mathbb E_{\mathrm{RQMC}}
 [{{e_{N,j}}}(A_{\bm{x}})^2
  \mid\mathcal F_{N,j-1}]=\operatorname{Var}_{\mathrm{RQMC}}
 \bigl(
 Q_{N,j}\one_{B_{N,j,\bm{x}}}
 \mid\mathcal F_{N,j-1}
 \bigr)\le
 c_{b,t,D}M_jN^{-1-1/D}.
\end{aligned}
\]
The uniform tube constants make the bound and its
cutoff independent of $\bm{x}$ and the realized old
array. This proves \eqref{eq:md-local-variance}.
\end{proof}

\subsection{Conditional orthant-to-HK transfer}
\label{app:md-hk}
Write $[\bm{0},\bm{y}]=\prod_{r=1}^p[0,y_r]$ and
$\|Z\|_{2\mid\mathcal G}:=(\mathbb E[|Z|^2\mid\mathcal G])^{1/2}$.
The following lemma is the conditional \(L^2\) analogue of the
left-continuous part of the proof of
\citet[Theorem~1]{aistleitner2015functions}.
That proof provides the signed-measure representation used below;
conditional Minkowski replaces the final pathwise discrepancy bound.

\begin{lemma}[Conditional orthant-to-HK transfer]
\label{lem:orthant-hk}
Let $\mu$ and $\nu$ be random Borel probability measures on $[0,1]^p$,
and let $\mathcal G$ be a sigma-field. Put $\Delta:=\mu-\nu$.
Suppose that, for every $\bm{y}\in[0,1]^p$,
$\|\Delta([\bm{0},\bm{y}])\|_{2\mid\mathcal G}\le C$ almost surely,
where $C\ge0$ is deterministic. For every deterministic,
coordinatewise left-continuous $g:[0,1]^p\to\mathbb R$ with finite
usual HK variation,
\begin{equation}\label{eq:orthant-hk-transfer}
 \|\Delta(g)\|_{2\mid\mathcal G}\le C V_{\rm HK}(g)
 \quad\hbox{almost surely}.
\end{equation}
\end{lemma}
\begin{proof}
Applied to $g-g(\bm{1})$, the signed-measure construction in the proof
of \citet[Theorem~1]{aistleitner2015functions} gives a deterministic
finite signed Borel measure $\widehat\nu_g$ on $[0,1]^p$ such that
\begin{equation}\label{eq:hk-measure-representation}
 g(\bm{z})-g(\bm{1})=\int_{[0,1]^p}
          \one_{[\bm{0},\bm{y}]}(\bm{z})\,\widehat\nu_g(\dd \bm{y}),\qquad
 |\widehat\nu_g|([0,1]^p)=V_{\rm HK}(g).
\end{equation}
Since $\Delta$ is the difference of two probability measures,
$\Delta([0,1]^p)=0$. Signed Fubini gives
\(
 \Delta(g)=\int_{[0,1]^p}\Delta([\bm{0},\bm{y}])\,
               \widehat\nu_g(\dd \bm{y}).
\)
Since
$(\omega,\bm{y})\mapsto
\Delta_\omega([\bm{0},\bm{y}])$
is jointly measurable and $|\widehat\nu_g|$ is deterministic and
finite, conditional Minkowski yields
\begin{equation}\label{eq:hk-conditional-minkowski}
 \|\Delta(g)\|_{2\mid\mathcal G}
 \le\int_{[0,1]^p}\|\Delta([\bm{0},\bm{y}])\|_{2\mid\mathcal G}
                  \,|\widehat\nu_g|(\dd \bm{y})
 \le CV_{\rm HK}(g).
\end{equation}
This completes the proof.
\end{proof}

\begin{remark}
Theorem~1 of \citet{aistleitner2015functions} itself does not require
left continuity.  Its proof removes that restriction by constructing
a left-continuous approximation adapted to the particular quadrature
points and then using a pathwise star-discrepancy bound.  Here we
control instead
\(
  \sup_y
  \|\Delta([0,y])\|_{2\mid\mathcal G},
\)
not the conditional \(L^2\) norm of the pathwise supremum.
That sample-dependent reduction therefore does not directly apply.
The one-dimensional \Cref{lem:bv-transfer} removes this restriction separately by
a deterministic Jordan-decomposition argument.
\end{remark}

For a deterministic bounded measurable $g$, define the innovation
$\eta_{N,s}(g):=\widehat\nu_{N,s}(g)-
\widehat\nu_{N,s-1}(\mathsf P_sg)$.
Fix $s$ and condition on $\mathcal F_{N,s-1}$. Define two random
probability measures on $[0,1]^p$ by
\[
 \mu^{\bm{\psi}}_{N,s}(C):=\widehat\nu_{N,s}(\bm{\psi}^{-1}(C)),\qquad
 \rho^{\bm{\psi}}_{N,s}(C):=
 (\widehat\nu_{N,s-1}\mathsf P_s)(\bm{\psi}^{-1}(C)),
\]
for Borel $C\subseteq[0,1]^p$. Fix $\bm{y}\in[0,1]^p$ and define
\(
 H_{\bm{y}}
 :=\{\bm{x}\in\mathcal X:
       \bm{\psi}(\bm{x})\in[\bm{0},\bm{y}]\}.
\)
By the definitions of the transformed measures and
\eqref{eq:md-generic-event},
\begin{equation}\label{eq:transformed-orthant-error}
\begin{aligned}
 \mu^{\bm{\psi}}_{N,s}([\bm{0},\bm{y}])
 -\rho^{\bm{\psi}}_{N,s}([\bm{0},\bm{y}])
 =
 \widehat\nu_{N,s}(H_{\bm{y}})
 -\widehat\nu_{N,s-1}
       (\mathsf P_s\one_{H_{\bm{y}}})={{e_{N,s}}}(H_{\bm{y}}).
\end{aligned}
\end{equation}

Since $\bm{\psi}$ is componentwise continuous and
strictly increasing, $H_{\bm{y}}$ is either empty
or a lower orthant, possibly with unrestricted
coordinates. If $H_{\bm{y}}=A_{\bm{x}}$ for some
$\bm{x}\in\mathbb R^p$, \Cref{lem:md-local} gives
\[
 \mathbb E_{\mathrm{RQMC}}
 [{{e_{N,s}}}(H_{\bm{y}})^2
  \mid\mathcal F_{N,s-1}]
 \le c_{b,t,D}M_sN^{-1-1/D},
 \qquad N\ge N_{0,s}.
\]

If $H_{\bm{y}}=\varnothing$, its local error is
zero. In the remaining cases, the product-interval
structure of $\mathcal X$ and the coordinatewise
monotonicity of $\bm{\psi}$ allow us to choose
deterministic finite thresholds
$\bm{x}^{(m)}\in\mathbb R^p$ such that $A_{\bm{x}^{(m)}}\uparrow H_{\bm{y}}$ as $m\to\infty$. Hence,
\(
 {{e_{N,s}}}(A_{\bm{x}^{(m)}})
 \longrightarrow
 {{e_{N,s}}}(H_{\bm{y}})
\) almost surely.
Applying conditional Fatou's lemma and \eqref{eq:transformed-orthant-error},
for every fixed $\bm{y}\in[0,1]^p$ and
admissible $N\ge N_{0,s}$,
\[
\begin{aligned}
 \bigl\|
 \mu^{\bm{\psi}}_{N,s}([\bm{0},\bm{y}])
 -\rho^{\bm{\psi}}_{N,s}([\bm{0},\bm{y}])
 \bigr\|_{2\mid\mathcal F_{N,s-1}}^2 &= \mathbb E_{\mathrm{RQMC}}
 [{{e_{N,s}}}(H_{\bm{y}})^2
  \mid\mathcal F_{N,s-1}]\\
 &\le
 \liminf_{m\to\infty}
 \mathbb E_{\mathrm{RQMC}}
 [{{e_{N,s}}}(A_{\bm{x}^{(m)}})^2
  \mid\mathcal F_{N,s-1}]\\ &\le c_{b,t,D}M_sN^{-1-1/D}.
\end{aligned}
\]
Moreover, the extension in \Cref{thm:md-fixed} satisfies
$\widetilde G_s\circ\bm{\psi}=G_s$ on $\mathcal X$, and hence
\[
 \mu^{\bm{\psi}}_{N,s}(\widetilde G_s)
 -\rho^{\bm{\psi}}_{N,s}(\widetilde G_s)
 =\widehat\nu_{N,s}(G_s)-\widehat\nu_{N,s-1}(\mathsf P_sG_s)
 =\eta_{N,s}(G_s).
\]
Applying \Cref{lem:orthant-hk} now yields
\begin{equation}\label{eq:md-hk-local}
 \mathbb E_{\mathrm{RQMC}}[
   \eta_{N,s}(G_s)^2\mid\mathcal F_{N,s-1}]
 \le c_{b,t,D}M_sV_{\rm HK}(\widetilde G_s)^2N^{-1-1/D}.
\end{equation}

\subsection{Innovation decomposition and proof of \texorpdfstring{\Cref{thm:md-fixed}}{Theorem \getrefnumber{thm:md-fixed}}}
\label{app:innovation-identities}
For a deterministic bounded measurable $g$, condition on
$\mathcal F_{N,s-1}$. Exact matching expresses $\widehat\nu_{N,s}(g)$
as the point-set average of
$g(\bm{\Psi}_s(\bm{Z}_{N,s-1}(v),\bm{u},w))$.
Its integral is $\widehat\nu_{N,s-1}(\mathsf P_sg)$, because each rank
interval has length $1/N$. Taking expectations over the original point
labels, as in \Cref{app:md-local}, gives
\begin{equation}\label{eq:innovation-center}
 \mathbb E_{\mathrm{RQMC}}[\widehat\nu_{N,s}(g)\mid\mathcal F_{N,s-1}]
 =\widehat\nu_{N,s-1}(\mathsf P_sg),\qquad
 \mathbb E_{\mathrm{RQMC}}[\eta_{N,s}(g)\mid\mathcal F_{N,s-1}]=0.
\end{equation}
Iteration from $\widehat\nu_{N,0}=\nu_0=\delta_{\bm{x}_0}$ proves unbiasedness.

{{Applying this to bounded nonnegative truncations and using monotone convergence extends one-time unbiasedness to nonnegative measurable tests. Positive and negative parts then give \eqref{eq:one-time-unbias} for every $f\in L^1(\nu_j)$.}}

Put $\Delta_{N,j}=\widehat\nu_{N,j}-\nu_j$. Since
$\nu_j=\nu_{j-1}\mathsf P_j$, the finite signed-measure recursion is
\[
 \Delta_{N,j}(g)=\eta_{N,j}(g)+\Delta_{N,j-1}(\mathsf P_jg),\qquad \Delta_{N,0}=0.
\]
Its finite expansion gives
\begin{equation}\label{eq:single-time-innovations}
 \Delta_{N,j}(f)=\sum_{s=1}^j\eta_{N,s}(\mathsf P_{s+1:j}f).
\end{equation}
Using $G_s$ from \eqref{eq:md-propagated} and interchanging finite sums,
\begin{equation}\label{eq:innovation-decomposition}
 \widehat\mu_{N,T}(f)-\mu_T(f)
 =\frac1L\sum_{j=T_0+1}^T\sum_{s=1}^j
        \eta_{N,s}(\mathsf P_{s+1:j}f)
 =\frac1L\sum_{s=1}^T\eta_{N,s}(G_s).
\end{equation}
Each $G_s$ is deterministic and bounded, so \eqref{eq:innovation-center}
also centers $\eta_{N,s}(G_s)$ conditionally.

For $s<t$, $\eta_{N,s}(G_s)$ is $\mathcal F_{N,t-1}$-measurable, and
\eqref{eq:innovation-center} implies
\[
 \mathbb E_{\mathrm{RQMC}}[\eta_{N,s}(G_s)\eta_{N,t}(G_t)]
 =\mathbb E_{\mathrm{RQMC}}\!\left[
    \eta_{N,s}(G_s)\,
    \mathbb E_{\mathrm{RQMC}}[\eta_{N,t}(G_t)\mid\mathcal F_{N,t-1}]
   \right]=0.
\]
All innovations are bounded and hence square integrable. Squaring
\eqref{eq:innovation-decomposition} therefore yields the general identity
\begin{equation}\label{eq:orthogonal-variance}
 \begin{aligned}
 \mathbb E_{\mathrm{RQMC}}[(\widehat\mu_{N,T}(f)-\mu_T(f))^2]
 =\operatorname{Var}_{\mathrm{RQMC}}(\widehat\mu_{N,T}(f))=\frac1{L^2}\sum_{s=1}^T
       \mathbb E_{\mathrm{RQMC}}[\eta_{N,s}(G_s)^2].
 \end{aligned}
\end{equation}
Applying \eqref{eq:md-hk-local} in \eqref{eq:orthogonal-variance}
and taking expectations gives
\[
 \mathbb E_{\mathrm{RQMC}}[(\widehat\mu_{N,T}(f)-\mu_T(f))^2]
 \le\frac{c_{b,t,D}}{L^2}\sum_{s=1}^T
       M_sV_{\rm HK}(\widetilde G_s)^2 N^{-1-1/D}.
\]
The maximum of the time-specific cutoffs in \eqref{eq:md-cutoff} ensures
every local bound holds. Taking $C_{b,t,D}=c_{b,t,D}$ gives
\eqref{eq:md-fixed-bound}.
This completes the proof of \Cref{thm:md-fixed}.

\section{Proof of the verified one-dimensional theorem}
\label{app:onedim}

\begin{proof}[Proof of \Cref{thm:main}]
By \Cref{thm:tube}, the event regions
$B_{N,s,x}$ in \eqref{eq:event} satisfy
\(
 \lambda_3((\partial B_{N,s,x})_\varepsilon)
 \le M_s\varepsilon,
\)
where $0<\varepsilon<1/2$ and $M_s=14C_{a,s}+6C_{\beta,s}+84$.
The bound is uniform over $x\in\mathbb R$,
admissible $N$, and realized old arrays.
Applying \Cref{lem:md-local}
with $D=3$ yields, for $N\ge N_0$,
\begin{equation}\label{eq:full-past-local}
\begin{aligned}
 \mathbb E_{\mathrm{RQMC}}
 [{{e_{N,s}}}(A_x)^2
  \mid\mathcal F_{N,s-1}]
 \le c_{b,t,3}M_sN^{-4/3}{{,}}
\end{aligned}
\end{equation}
uniformly over $x\in\mathbb R$, where $c_{b,t,3}$ is defined in
\eqref{eq:tube-variance-constant} and the cutoff $N_0$ is given by
\begin{equation}\label{eq:main-cutoff}
 k_0:=\max\{0,\lfloor\log_b(2\sqrt3)\rfloor+1\},
 \qquad N_0:=b^{t+3k_0}.
\end{equation}

Apply the conditional pointwise-BV transfer
of \Cref{lem:bv-transfer} to
$\widehat\nu_{N,s}
-\widehat\nu_{N,s-1}\mathsf P_s$.
For every deterministic pointwise
$g\in\mathrm{BV}(\mathcal X)$,
\begin{equation}\label{eq:local-bv-innovation}
 \mathbb E_{\mathrm{RQMC}}
 [\eta_{N,s}(g)^2
  \mid\mathcal F_{N,s-1}]
 \le c_{b,t,3}M_s
      V_{\mathcal X}(g)^2N^{-4/3}.
\end{equation}

The finite transition-CDF variation
constants $\Lambda_j$ and
\Cref{lem:bv-transfer} imply
\eqref{eq:Gs-bv-bound}:
\[
 V_{\mathcal X}(G_s)
 \le\mathcal H_s^{T_0,T}
      V_{\mathcal X}(f).
\]
Thus \eqref{eq:local-bv-innovation}
applies with $g=G_s$. Finally, the innovation decomposition and
orthogonality identity
\eqref{eq:orthogonal-variance} yield
\begin{equation}\label{eq:innovation-variance}
\begin{aligned}
 \mathbb E_{\mathrm{RQMC}}
 [(\widehat\mu_{N,T}(f)-\mu_T(f))^2]
 &=
 \frac1{L^2}\sum_{s=1}^T
 \mathbb E_{\mathrm{RQMC}}
 [\eta_{N,s}(G_s)^2]\\
 &\le
 \frac{c_{b,t,3}N^{-4/3}}{L^2}
 \sum_{s=1}^T M_sV_{\mathcal X}(G_s)^2\\
 &\le
 \frac{c_{b,t,3}V_{\mathcal X}(f)^2}{L^2}
 \sum_{s=1}^T
 M_s(\mathcal H_s^{T_0,T})^2
 N^{-4/3}.
\end{aligned}
\end{equation}
This completes the proof.
\end{proof}

\clearpage  

\setcounter{section}{0}
\setcounter{page}{1}
\setcounter{figure}{0}
\setcounter{table}{0}

\renewcommand{\thesection}{SM\arabic{section}}
\renewcommand{\thepage}{SM\arabic{page}}
\renewcommand{\thefigure}{SM\arabic{figure}}
\renewcommand{\thetable}{SM\arabic{table}}

\renewcommand{\theHsection}{SM.\arabic{section}}
\renewcommand{\theHsubsection}{\theHsection.\arabic{subsection}}
\renewcommand{\theHsubsubsection}{\theHsubsection.\arabic{subsubsection}}
\renewcommand{\theHequation}{\theHsection.\arabic{equation}}
\renewcommand{\theHtheorem}{\theHsection.\arabic{theorem}}
\renewcommand{\theHfigure}{SM.\arabic{figure}}
\renewcommand{\theHtable}{SM.\arabic{table}}
\renewcommand{\theHalgorithm}{\theHsection.\arabic{algorithm}}

\crefalias{section}{section}

\section*{Supplementary Materials: Fixed-Horizon Error Rates for Array-RQMC Metropolis--Hastings}

\Cref{sec:supp-supporting} proves the pointwise BV transfer and uniform
MH boundary-tube lemmas used in \Cref{thm:main} of the main paper.
\Cref{sec:supp-variation} proves
\Cref{prop:kernel-variation,prop:symmetric-location}, and
\Cref{app:targets} verifies the target classes in \Cref{tab:state-variation},
including parameter-uniform mixture bounds at fixed component count.
\Cref{sec:supp-product-mwg} proves the product-target MWG reduction
and its additive polynomial extension, with the scalar polynomial
estimate proved in \Cref{sec:supp-polynomial}.
\Cref{app:inversion} compares the framework with one-dimensional inversion.
\Cref{sec:supp-horizon} places the obstruction to whole-line BV contraction
alongside the numerical horizon experiment.

\section{Supporting lemmas for the one-dimensional fixed-horizon theorem}
\label{sec:supp-supporting}
The two lemmas below supply the local geometry and pointwise BV control
used in \Cref{thm:main}; their statements appear in
\Cref{lem:bv-transfer,thm:tube} of the main paper.

\subsection{Pointwise BV transfer and propagation}
\label{sec:supp-bv-transfer}
\begin{proof}[Proof of \Cref{lem:bv-transfer}]
    Extend measures by zero outside $\mathcal X$. Extend $f$ constantly beyond
finite endpoints, using its one-sided limit at an excluded endpoint and
its actual value at an included endpoint. This preserves variation and
all integrals. A singleton interval is immediate. On $\mathbb R$, let
$f_{\pm}=\lim_{x\to{\pm}\infty}f(x)$ and $v(x)=V_{(-\infty,x]}(f)$.
For $x<y$, $v(y)-v(x)\ge|f(y)-f(x)|$. Thus
\(
 {{f_1}}=(v+f-f_-)/2,\ {{f_2}}=(v-f+f_-)/2
\)
are bounded nondecreasing functions, $f=f_-+{{f_1-f_2}}$, and their range lengths
sum to $V_{\mathbb R}(f)$. For either such function {{$g\in\{f_1,f_2\}$, with limits $g_\pm$ at $\pm\infty$}},
    \(
 {{g(x)= g_-+\int_{g_-}^{g_+}\one\{g(x)>a\}}}\,\dd a.
\)
Every superlevel set here is an open or closed upper half-line, or is
empty or the whole line. The corresponding lower half-lines are controlled:
open lower half-lines are increasing limits of closed ones, and {{the bound}} $|\Delta(A)|\le 1${{, valid for every measurable $A$,}}
justifies $L^2$ convergence (or conditional Fatou). Since $\Delta(\mathcal X)=0$,
complements have the same bound. Minkowski's integral inequality, or
conditional Cauchy--Schwarz and Fubini, therefore gives
{$$
\|\Delta(f)\|_2\le
{{\|\Delta(f_1)\|_2+\|\Delta(f_2)\|_2}}
\le C{{\{(f_{1,+}-f_{1,-})+(f_{2,+}-f_{2,-})\}}} = CV_{\mathbb R}(f).
$$}
The conditional proof uses the bound for each fixed level set and Fubini.

For the operator assertion, open lower-half-line probabilities are
pointwise limits of $K(x\mid z)$ as the threshold increases to its endpoint.
Their state variation is at most $\Lambda$, by passing to the limit on
any fixed finite state partition. Complements have the same variation.
Apply the displayed level-set representation, sum absolute increments
on a state partition under the integral, and then take the supremum.
This proves \eqref{eq:kernel-bv-operator}. The argument preserves the
actual values of $f$ on $\mathcal X$, including at atoms.
\end{proof}

\subsection{Uniform MH boundary-tube bound}
\label{sec:supp-boundary}
\label{app:boundary}
We prove \Cref{thm:tube} for the scalar event \eqref{eq:event}
defined in \Cref{sec:general-main} of the main paper. Write $\beta_{N,j,x}(v)=\beta_{j,x}(Z_{N,j-1}(v))$.

In $\mathbb U=[0,1)^3$, let $E_{a,j}$ be the acceptance event
$\{(v,u,w)\in\mathbb U:u>0,\ w\le a_{N,j}(v,u)\}$,
let $E_{\beta,j}=\{(v,u,w)\in\mathbb U:u\le\beta_{j,x}(Z_{N,j-1}(v))\}$
be the proposal-threshold event, and let
$E_{x,j}=\{(v,u,w)\in\mathbb U:Z_{N,j-1}(v)\le x\}$
be the current-state threshold event. The generalized inverse identity
and the convention $\Psi_j(z,0,w)=z$ give the exact MH decomposition
\begin{equation}\label{eq:actual-to-proxy-boundary}
 B_{N,j,x}=(E_{a,j}\cap E_{\beta,j})
 \cup\bigl((\mathbb U\setminus E_{a,j})\cap E_{x,j}\bigr).
\end{equation}

The decomposition in \eqref{eq:actual-to-proxy-boundary} reduces the MH boundary to finitely many simpler surfaces.
For any sets \(G,H\subset\mathbb R^D\), the boundaries of \(G\cup H\),
\(G\cap H\), and \(G\setminus H\) lie in \(\partial G\cup\partial H\);
for \(E\subset\mathbb U\), its boundary lies in
\((\partial E\cap(0,1)^3)\cup\partial\mathbb U\). Hence
\eqref{eq:actual-to-proxy-boundary} gives
\begin{shrinkeq}{0em}
\begin{equation}\label{eq:boundary-union-global}
\partial B_{N,j,x}
\subset (\partial E_{a,j}\cap(0,1)^3)
\cup(\partial E_{\beta,j}\cap(0,1)^3)
\cup(\partial E_{x,j}\cap(0,1)^3)
\cup\partial\mathbb U.
\end{equation}
\end{shrinkeq}
A Euclidean \(\varepsilon\)-tube lies in the corresponding supremum norm tube, and
the tube of a finite union lies in the union of its tubes. Therefore
\begin{shrinkeq}{0em}
\begin{equation}\label{eq:MH-tube-decomposition}
\begin{aligned}
\lambda_3((\partial B_{N,j,x})_\varepsilon)
&\le \lambda_3(((\partial E_{a,j}\cap(0,1)^3)_\varepsilon^{(\infty)}))
 +\lambda_3(((\partial E_{\beta,j}\cap(0,1)^3)_\varepsilon^{(\infty)}))\\
&\quad+\lambda_3(((\partial E_{x,j}\cap(0,1)^3)_\varepsilon^{(\infty)}))
 +\lambda_3((\partial\mathbb U)_\varepsilon^{(\infty)}){{,}}
\end{aligned}
\end{equation}
\end{shrinkeq}
where ${{\Gamma}}_\varepsilon^{(\infty)}
:=\{\bm{x}:\inf_{\bm{y}\in{{\Gamma}}}\|\bm{x}-\bm{y}\|_\infty\le\varepsilon\}$ {{for $\Gamma\subset\mathbb R^3$}}.
Consider the terms on the right-hand side separately for \(0<\varepsilon<1/2\).
\begin{itemize}
    \item
\begin{shrinkeq}{0em}
\begin{equation}\label{eq:l3u}
    \begin{aligned}
\lambda_3\!\left(
(\partial\mathbb U)_{\varepsilon}^{(\infty)}
\right)
&=
(1+2\varepsilon)^3
-
(1-2\varepsilon)^3 =
12\varepsilon+16\varepsilon^3 < 16\varepsilon.
\end{aligned}
\end{equation}
\end{shrinkeq}
\item Let $r_{x}:=\#\{i\in\{0,\ldots,N-1\}:z_i\le x\}.$ Since $Z_{N,j-1}$ is nondecreasing, we have
\begin{shrinkeq}{0em}
\[
E_{x,j}
=
[0,r_{x}/N)\times [0,1)^2 \implies \partial E_{x,j}\cap(0,1)^3
  \subset
  \{r_x/N\}\times[0,1]^2.
\]
\end{shrinkeq}
Therefore,
\begin{shrinkeq}{0em}
\begin{equation}
  \lambda_3
  \left(
    (
      \partial E_{x,j}\cap(0,1)^3
    )_\varepsilon^{(\infty)}
  \right)
  \le
  2\varepsilon(1+2\varepsilon)^2
  <
  8\varepsilon.
  \label{eq:l3x}
\end{equation}
\end{shrinkeq}
\item Write
\(D_{\beta,j}:=\{(v,u)\in[0,1)^2:u\le\beta_{N,j,x}(v)\}\). Since
\(E_{\beta,j}=D_{\beta,j}\times[0,1)\), the standard boundary identity for products
restricted to the open cube gives
\begin{shrinkeq}{0em}
\[
\partial E_{\beta,j}\cap(0,1)^3
=(\partial D_{\beta,j}\cap(0,1)^2)\times(0,1).
\]
\end{shrinkeq}
Consequently,
\begin{shrinkeq}{0em}
\begin{equation}\label{eq:Ebeta-boundary-product}
\begin{aligned}
\lambda_3(((\partial E_{\beta,j}\cap(0,1)^3)_\varepsilon^{(\infty)}))
&=(1+2\varepsilon)
\lambda_2(((\partial D_{\beta,j}\cap(0,1)^2)_\varepsilon^{(\infty)}))\\
&\le2\lambda_2(((\partial D_{\beta,j}\cap(0,1)^2)_\varepsilon^{(\infty)})).
\end{aligned}
\end{equation}
\end{shrinkeq}
\end{itemize}
The remaining acceptance and proposal-threshold boundaries,
\(\partial E_{a,j}\) and \(\partial D_{\beta,j}\), are controlled by the
next two lemmas.

\begin{lemma}[Tube bound for oscillation subgraphs]
\label{lem:oscillation-subgraph}
Let $h:[0,1)\times(0,1)\to[0,1]$ be {{a function}}. Suppose that
\begin{shrinkeq}{0em}
\begin{equation}
 \sup_{k\geq 1} \frac{1}{k}\sum_{\jmath ,q=0}^{k-1}
 \operatorname{osc}_{J_\jmath ^{(k)}\times K_q^{(k)}}(h)
 \le C_h < \infty.
 \label{eq:h-grid-osc}
\end{equation}
\end{shrinkeq}
Define $G(h):=\{(v,u,w)\in[0,1)^3: u > 0,w\leq h(v,u)\}$.
Then, for every $0<\varepsilon<1/2$,
\begin{shrinkeq}{0em}
\begin{equation*}
 \lambda_3((\partial G(h)
      \cap (0,1)^3)^{(\infty)}_\varepsilon)
 <14(C_h+3)\varepsilon.
\end{equation*}
\end{shrinkeq}
\end{lemma}
\begin{proof}
Fix $0<\eps<1/2$ and set $k=\left\lceil\eps^{-1}\right\rceil$.
Then $1/k\leq\eps$ and $k\leq\eps^{-1}+1<2\eps^{-1}$. For \(\jmath ,q=0,\ldots,k-1\), set $Q_{\jmath q}=J_\jmath ^{(k)}\times K_q^{(k)}$.
For every $Q_{\jmath q}$, define
\begin{shrinkeq}{0em}
\begin{equation*}
\omega_{\jmath q} = \operatorname{osc}_{Q_{\jmath q}}(h)
=
\overline h_{\jmath q}-\underline h_{\jmath q}, \qquad \underline h_{\jmath q}
:=
\inf_{(v,u)\in Q_{\jmath q}}h(v,u),
\qquad
\overline h_{\jmath q}
:=
\sup_{(v,u)\in Q_{\jmath q}}h(v,u).
\end{equation*}
\end{shrinkeq}
Define the corresponding vertical prism $S_{\jmath q}
:=
Q_{\jmath q}\times[\underline h_{\jmath q},\overline h_{\jmath q}].$ We first prove the boundary covering
\begin{shrinkeq}{0em}
\begin{equation}
\label{eq:boundary_localization}
 \partial G(h) \cap (0,1)^3
\subset
\bigcup_{\jmath ,q=0}^{k-1}S_{\jmath q}.
\end{equation}
\end{shrinkeq}
Take $\bm{x}=(v,u,w)\in \partial G(h) \cap (0,1)^3$ and $\bm{y}=(v,u) \in (0,1)^2$. Since \(\bm{x}\in(0,1)^3\), there exists a constant $\rho_{\bm{x}}>0$ such that $B_\infty(\bm{x},\rho_{\bm{x}})\subset(0,1)^3.$
Define the grid star of $\bm{y}$ by
\begin{shrinkeq}{0em}
\[
\operatorname{St}(\bm{y})
=
\bigcup_{(\jmath ,q)\in\mathcal {I}(\bm{y})}Q_{\jmath q}, \quad \mathcal {I}(\bm{y}):=\{(\jmath ,q):\bm{y}\in Q_{\jmath q}\}.
\]
\end{shrinkeq}
Because the adjacent grid cells overlap at their common endpoints, \(\operatorname{St}(\bm{y})\) contains a neighborhood
of \(\bm{y}\). Therefore there exists $\rho_0>0$ such that $B_\infty(\bm{y},\rho_0) \subset
\operatorname{St}(\bm{y}).$

If
$w\in[\underline h_{\jmath q},\overline h_{\jmath q}]$ for some
$(\jmath ,q)\in\mathcal {I}(\bm{y})$, then $\bm{x}\in S_{\jmath q}$. Suppose, to the contrary, that
\begin{shrinkeq}{0em}
\[
w\notin
[\underline h_{\jmath q},\overline h_{\jmath q}]
\qquad
\text{for every }(\jmath ,q)\in\mathcal I(\bm{y}).
\]
\end{shrinkeq}
Since $\bm{y}\in Q_{\jmath q}$ for every $(\jmath ,q)\in\mathcal I(\bm{y})$, $\underline h_{\jmath q}
\leq h(\bm{y})
\leq\overline h_{\jmath q}$. Thus $w\neq h(\bm{y})$.

Assume first that $w<h(\bm{y})$. It follows that $w<\underline h_{\jmath q}$ for every $(\jmath ,q)\in\mathcal {I}(\bm{y}).$
Since $\mathcal {I}(\bm{y})$ is finite,
\begin{shrinkeq}{0em}
\[
\delta
:=
\min_{(\jmath ,q)\in\mathcal {I}(\bm{y})}
(\underline h_{\jmath q}-w)
>0.
\]
\end{shrinkeq}
Choose $0<\rho<\min\{\rho_0,\delta,\rho_{\bm{x}}\}.$
For any $\bm{x}_0=(\bm{y}_0,w_0)\in B_\infty(\bm{x},\rho) \subset (0,1)^3$, since
$\bm{y}_0\in B_\infty(\bm{y},\rho) \subset\operatorname{St}(\bm{y})$, $\bm{y}_0\in Q_{\jmath q}$ for some $(\jmath ,q)\in\mathcal {I}(\bm{y})$,
\begin{shrinkeq}{0em}
\[
h(\bm{y}_0)
\geq\underline h_{\jmath q}
\geq w+\delta > w+\rho > w_0.
\]
\end{shrinkeq}
Hence, $\bm{x}_0\in G(h)$. Thus $B_\infty(\bm{x},\rho)\subset G(h)$, so \(\bm{x}\) is an interior point of \(G(h)\), contradicting
\(\bm{x}\in\partial G(h)\). The case \(w>h(\bm{y})\) is analogous.
Let $\delta_1
:=
\min_{(\jmath ,q)\in\mathcal {I}(\bm{y})}
(w-\overline h_{\jmath q})
>0$
and choose $0<\rho_1<\min\{\rho_0,\delta_1,\rho_{\bm{x}}\}$. For any $\bm{x}_1=(\bm{y}_1,w_1)\in B_\infty(\bm{x},\rho_1)\subset (0,1)^3$, $\bm{x}_1 \notin  G(h)$, again contradicting \(\bm{x}\in\partial G(h)\). This proves \eqref{eq:boundary_localization}. Consequently, by $1/k\leq\eps$, we obtain
\begin{shrinkeq}{0em}
\begin{equation}
    \begin{aligned}
\lambda_3(( \partial G(h) \cap (0,1)^3)_\eps^{(\infty)})
&\leq
\sum_{\jmath ,q}
\lambda_3((S_{\jmath q})_\eps^{(\infty)}) = \sum_{\jmath ,q}(1/k+2\eps)^2(\omega_{\jmath q}+2\eps) \\
& \le
9\eps^2\sum_{\jmath ,q}\omega_{\jmath q}
+
18\eps^3k^2 \le 9\eps^2C_hk+18\eps^3k^2\\
& = 9\eps k(C_h+2\eps k)\eps < 14(C_h+3)\eps.
\label{eq:tube_reduced}
\end{aligned}
\end{equation}
\end{shrinkeq}
Here we use $\eps k<3/2$ and \eqref{eq:h-grid-osc}. This completes the proof.
\end{proof}

\begin{lemma}[Tube bound for BV threshold subgraphs]
\label{lem:threshold-cylinder}
Let $\zeta:[0,1)\to[0,1]$ be a function of bounded variation
 satisfying $V_{[0,1)}(\zeta)\leq B.$
 Define $\mathcal S_\zeta:=\{(v,u)\in[0,1)^2: u\leq \zeta(v)\}$.
Then, for every $0<\varepsilon<1/2$,
\begin{shrinkeq}{0em}
\begin{equation*}
 \lambda_2((\partial\mathcal S_\zeta \cap (0,1)^2)_\varepsilon^{(\infty)})
 < 3(B+3)\varepsilon.
\end{equation*}
\end{shrinkeq}
\end{lemma}

\begin{proof}
Fix \(0<\varepsilon<1/2\), set \(k=\lceil\varepsilon^{-1}\rceil\), and for each
\(\jmath \) define
\begin{shrinkeq}{0em}
\[
\underline\zeta_\jmath :=\inf_{J_\jmath ^{(k)}}\zeta,
\qquad \overline\zeta_\jmath :=\sup_{J_\jmath ^{(k)}}\zeta,
\qquad w_\jmath :=\overline\zeta_\jmath -\underline\zeta_\jmath ,
\qquad R_\jmath :=J_\jmath ^{(k)}\times[\underline\zeta_\jmath ,\overline\zeta_\jmath ].
\]
\end{shrinkeq}
The one-dimensional analogue of the boundary covering argument in
\Cref{lem:oscillation-subgraph} gives
\(\partial\mathcal S_\zeta\cap(0,1)^2\subset\bigcup_{\jmath =0}^{k-1}R_\jmath \). Hence
\begin{shrinkeq}{0em}
\[
\begin{aligned}
\lambda_2((\partial\mathcal S_\zeta\cap(0,1)^2)_\varepsilon^{(\infty)})
&\le\sum_{\jmath =0}^{k-1}(1/k+2\varepsilon)(w_\jmath +2\varepsilon)\\
&\le(1/k+2\varepsilon)(B+2k\varepsilon)
<3(B+3)\varepsilon,
\end{aligned}
\]
\end{shrinkeq}
because \(1/k\le\varepsilon\), \(k\varepsilon<3/2\), and
\(\sum_\jmath w_\jmath \le V_{[0,1)}(\zeta)\le B\).
\end{proof}

We can now combine these estimates to establish the uniform
boundary-tube bound \eqref{eq:B-tube-bound}.

\begin{proof}[Proof of \Cref{thm:tube}]
The finite quantity $C_{a,j}$ in \eqref{eq:general-quantities} gives
\eqref{eq:h-grid-osc} for every sorted array with $C_h=C_{a,j}$.
Applying \Cref{lem:oscillation-subgraph}, we have
\begin{shrinkeq}{0em}
\begin{equation}
  \lambda_3
  \left(
    \left(
      \partial E_{a,j}
      \cap (0,1)^3
    \right)_\varepsilon^{(\infty)}
  \right)
  <
  14(C_{a,j}+3)\varepsilon.
  \label{eq:Ea-tube-global}
\end{equation}
\end{shrinkeq}
For the proposal threshold set, since $\beta_{N,j,x}(v)=\beta_{j,x}(Z_{N,j-1}(v))$ and $Z_{N,j-1}$ is nondecreasing, composition with $Z_{N,j-1}$ does not increase
one-dimensional total variation:
\begin{shrinkeq}{0em}
\begin{equation*}
 V_{[0,1)}( \beta_{N,j,x})
 \leq V_{\mathcal X}(\beta_{j,x})
 \leq C_{\beta,j}.
\end{equation*}
\end{shrinkeq}
Applying \Cref{lem:threshold-cylinder} therefore gives
\begin{shrinkeq}{0em}
\begin{equation}
 \lambda_2
  \left((\partial D_{\beta,j}\cap (0,1)^2)_\varepsilon^{(\infty)}\right)
 < 3(C_{\beta,j}+3)\varepsilon.
 \label{eq:Ebeta-tube-global}
\end{equation}
\end{shrinkeq}
Combining \eqref{eq:MH-tube-decomposition}, \eqref{eq:l3u}, \eqref{eq:l3x}, \eqref{eq:Ebeta-boundary-product}, \eqref{eq:Ea-tube-global}, together with \eqref{eq:Ebeta-tube-global} yields
\begin{shrinkeq}{0em}
\[
 \lambda_3((\partial B_{N,j,x})_\varepsilon) < 14(C_{a,j}+3)\varepsilon + 6(C_{\beta,j}+3)\eps + 16\eps+8\eps =
 (14C_{a,j}+6C_{\beta,j}+84)\varepsilon.
\]
\end{shrinkeq}
Only $C_{a,j}$ and $C_{\beta,j}$ enter this estimate, so it is uniform
over all $N$, thresholds and sorted arrays in the state interval.
\end{proof}

\section{Verification tools for symmetric location proposals}
\label{sec:supp-variation}
We first bound transition-CDF variation and then use that bound to verify
the quantitative conditions for symmetric location proposals.

\subsection{Transition-CDF variation}
\label{app:kernel}
We prove \Cref{prop:kernel-variation} using a moving-limit estimate that
accounts for acceptance and rejection separately.

\begin{lemma}[Variation of integrals with moving limits]
\label{lem:moving-limit-variation}
Let \(I\subseteq\mathbb R\) be an interval, let
\(\beta:I\to[0,1]\) be of bounded variation, and let
\(a:I\times(0,1)\to[0,1]\) be measurable. Define
\begin{shrinkeq}{0em}
\[
  A_I(a)
  :=
  \int_0^1 V_I\bigl(a(\cdot,u)\bigr)\,du, \qquad H_-(z)
  :=
  \int_0^{\beta(z)}a(z,u)\,du,
  \qquad
  H_+(z)
  :=
  \int_{\beta(z)}^1a(z,u)\,du.
\]
\end{shrinkeq}
Suppose that \(A_I(a)<\infty\). Then
\begin{shrinkeq}{0em}
\[
  V_I(H_-)
  \le
  A_I(a)+V_I(\beta),
  \qquad
  V_I(H_+)
  \le
  A_I(a)+V_I(\beta).
\]
\end{shrinkeq}
\end{lemma}

\begin{proof}
Let $z_0<z_1<\cdots<z_\ell$
be an arbitrary finite partition of \(I\). For \(H_-\), since \(0\le a\le1\), we have
\begin{shrinkeq}{0em}
\[
\begin{aligned}
\left|
  H_-(z_i)-H_-(z_{i-1})
\right|
&\le
\left|\int_0^{\beta(z_i)}
  \bigl[
    a(z_i,u)-a(z_{i-1},u)
  \bigr]\,du\right|+
\left|\int_{\beta(z_{i-1})}^{\beta(z_i)}a(z_{i-1},u)\,\dd u\right|\\
&\le
\int_0^1
\left|
  a(z_i,u)-a(z_{i-1},u)
\right|\,du+
\left|
  \beta(z_i)-\beta(z_{i-1})
\right|.
\end{aligned}
\]
\end{shrinkeq}
Summing over \(i\) and using linearity of the integral gives
\begin{shrinkeq}{0em}
\[
\begin{aligned}
\sum_{i=1}^{\ell}
\left|
  H_-(z_i)-H_-(z_{i-1})
\right|
&\le
\int_0^1
\sum_{i=1}^{\ell}
\left|
  a(z_i,u)-a(z_{i-1},u)
\right|\,du +
\sum_{i=1}^{\ell}
\left|
  \beta(z_i)-\beta(z_{i-1})
\right| \\
&\le
\int_0^1
V_I\bigl(a(\cdot,u)\bigr)\,du
+
V_I(\beta) = A_I(a) + V_I(\beta).
\end{aligned}
\]
\end{shrinkeq}
Taking the supremum over all finite partitions proves the bound for
\(H_-\). The analogous argument for \(H_+\) gives the second bound.
\end{proof}

\begin{proof}[Proof of \Cref{prop:kernel-variation}]
Write $A_j:=\int_0^1V_{\mathcal X}(a_j(\cdot,u))\,\dd u$.
Fix \(x\in\mathcal X\). By the generalized inverse construction, for any \(z\in\mathcal X\) and \(u\in(0,1)\),
\begin{shrinkeq}{0em}
\[
  Y_j(z,u)\le x
  \quad\Longleftrightarrow\quad
  u\le\beta_{j,x}(z).
\]
\end{shrinkeq}

If \(z>x\), rejection leaves the chain at the state \(z>x\).
Hence the final state is at most \(x\) precisely when the proposal is
both below \(x\) and accepted. Therefore,
\begin{shrinkeq}{0em}
\[
\begin{aligned}
K_j(x\mid z)
&=
\int_{(0,1)^2}
\mathbf 1
\left\{
  u\le\beta_{j,x}(z),
  \,
  w\le a_j(z,u)
\right\}
\,du\,dw\\
&=
\int_0^{\beta_{j,x}(z)}
a_j(z,u)\,du := H_{-}(z), \qquad z>x.
\end{aligned}
\]
\end{shrinkeq}

If \(z\le x\), rejection leaves the chain at or below \(x\). The
final state exceeds \(x\) precisely when the proposal is above \(x\)
and is accepted. Thus
\begin{shrinkeq}{0em}
\[
\begin{aligned}
1-K_j(x\mid z)
&=
\int_{(0,1)^2}
\mathbf 1
\left\{
  u>\beta_{j,x}(z),
  \,
  w\le a_j(z,u)
\right\}
\,du\,dw\\
&=
\int_{\beta_{j,x}(z)}^1
a_j(z,u)\,du= H_{+}(z), \qquad z\le x.
\end{aligned}
\]
\end{shrinkeq}
Combining the two cases gives
\begin{shrinkeq}{0em}
\[
  K_j(x\mid z)
  = \begin{cases}
      1-H_{+}(z), & z\in I_x^-:=\mathcal X\cap(-\infty,x],\\
      H_{-}(z), & z\in I_x^+:=\mathcal X\cap(x,\infty).
  \end{cases}
\]
\end{shrinkeq}
Applying \Cref{lem:moving-limit-variation} on the two subintervals gives
\begin{shrinkeq}{0em}
\[
\begin{aligned}
V_{I_x^-}\bigl(K_j(x\mid\cdot)\bigr) = V_{I_x^-}\bigl(1-H_{+}(z)\bigr) = V_{I_x^-}\bigl(
  H_{+}(z)\bigr)
&\le
\int_0^1
V_{I_x^-}\bigl(a_j(\cdot,u)\bigr)\,du
+
V_{I_x^-}(\beta_{j,x}),
\\
V_{I_x^+}\bigl(K_j(x\mid\cdot)\bigr) = V_{I_x^+}\bigl(H_{-}(z)\bigr)
&\le
\int_0^1
V_{I_x^+}\bigl(a_j(\cdot,u)\bigr)\,du
+
V_{I_x^+}(\beta_{j,x}).
\end{aligned}
\]
\end{shrinkeq}
For any $[0,1]$-valued function \(g\) on \(\mathcal X\), we have
\begin{shrinkeq}{0em}
\[
  V_{I_x^-}(g)+V_{I_x^+}(g)
  \le
  V_{\mathcal X}(g) \le V_{I_x^-}(g)+V_{I_x^+}(g) +1.
\]
\end{shrinkeq}
Consequently,
\begin{shrinkeq}{0em}
\[
\begin{aligned}
V_{\mathcal X}\bigl(K_j(x\mid\cdot)\bigr)
&\le
1
+
V_{I_x^-}\bigl(K_j(x\mid\cdot)\bigr)
+
V_{I_x^+}\bigl(K_j(x\mid\cdot)\bigr)
\\
&\le
1+ \int_0^1
V_{I_x^-}\bigl(a_j(\cdot,u)\bigr)\,du
+
V_{I_x^-}(\beta_{j,x}) + \int_0^1
V_{I_x^+}\bigl(a_j(\cdot,u)\bigr)\,du
+
V_{I_x^+}(\beta_{j,x})\\
& \le 1+\int_0^1 V_{\mathcal{X}}\bigl(a_j(\cdot,u)\bigr)\,du + V_{\mathcal X}(\beta_{j,x})
\le
1+A_j+C_{\beta,j}.
\end{aligned}
\]
\end{shrinkeq}
Taking the supremum over \(x\in\mathcal X\) proves \eqref{eq:global-Lambda}.
Thresholds outside the state interval give the constant CDF zero or one,
and excluded finite endpoints are also obtained as limits. Consequently,
the same $\Lambda_j$ controls all real thresholds, as required for
\Cref{lem:bv-transfer}.
\end{proof}

\subsection{Symmetric location proposals}
\label{sec:supp-symmetric}
\label{app:symmetric}
\begin{proof}[Proof of \Cref{prop:symmetric-location}]
Under the target and proposal hypotheses of \Cref{prop:symmetric-location},
we first prove
\begin{equation}\label{eq:acceptance-constants}
C_{a,j}\le B_\pi+4R+2,
\end{equation}
with the grid-oscillation bound uniform over $N$ and every realized
sorted array.

The location proposal \(Y_j(z,u)=z+\sigma_jH^{-1}(u)\) has a positive finite symmetric density, so
\begin{shrinkeq}{0em}
\begin{equation}\label{eq:gaussian-rw-acceptance}
  a_j(z,u)
  =
  1\wedge\frac{\pi(z+\sigma_jH^{-1}(u))}{\pi(z)},
  \qquad z\in\mathbb R,\quad u\in(0,1).
\end{equation}
\end{shrinkeq}
Fix \(N\ge1\), a sorted array, and \(k\ge1\). Write
\(Q_{\jmath q}=J_\jmath ^{(k)}\times K_q^{(k)}\) and
\(\omega_{\jmath q}=\operatorname{osc}_{Q_{\jmath q}}(a_{N,j})\).

For \(r=1,\ldots,R\), let
\(\eta_r(z)=H((c_r-z)/\sigma_j)\). Each \(\eta_r\) is strictly
decreasing from \(1\) to \(0\), so \(V_{\mathbb R}(\eta_r)=1\).
With \(\eta_0=0\) and \(\eta_{R+1}=1\),
\begin{shrinkeq}{0em}
\begin{equation}\label{eq:moving-strip-gaussian}
  \eta_\ell(z)<u<\eta_{\ell+1}(z)
  \quad\Longleftrightarrow\quad
  c_\ell<Y_j(z,u)<c_{\ell+1},
  \qquad \ell=0,\ldots,R.
\end{equation}
\end{shrinkeq}
For fixed \(z\), the denominator \(\pi(z)\) in
\eqref{eq:gaussian-rw-acceptance} is constant,
\(u\mapsto Y_j(z,u)\) is strictly increasing, and
\(x\mapsto1\wedge x\) is nondecreasing.
Since \(\pi\) has one fixed monotonicity direction on
\((c_\ell,c_{\ell+1})\), it follows from
\eqref{eq:moving-strip-gaussian} that \(u\mapsto a_j(z,u)\)
is monotone between the two moving boundaries
\(\eta_\ell(z)\) and \(\eta_{\ell+1}(z)\), with a direction depending
only on \(\ell\), not on \(z\).

\textbf{Bounding bad cells.}
For every \(r=1,\ldots,R\), define
\(\zeta_{r,N}(v):=\eta_r(Z_{N,j-1}(v))\), and set
\(\zeta_{0,N}\equiv0\) and \(\zeta_{R+1,N}\equiv1\).
Since \(Z_{N,j-1}\) is nondecreasing, each \(\zeta_{r,N}\) is
nonincreasing. For fixed \(\jmath \) and \(r\), define
$\underline\zeta_{\jmath ,r}
  :=
  \inf_{v\in J_\jmath ^{(k)}}\zeta_{r,N}(v)$, $\overline\zeta_{\jmath ,r}
  :=
  \sup_{v\in J_\jmath ^{(k)}}\zeta_{r,N}(v)$
and let $\mathcal B_{\jmath ,r}
  :=
  \left\{
    q\in\{0,\ldots,k-1\}:
    K_q^{(k)}
    \cap
    [\underline\zeta_{\jmath ,r},\overline\zeta_{\jmath ,r}]
    \ne\varnothing
  \right\}.$
Thus \(q\in\mathcal B_{\jmath ,r}\) whenever the {{\(r\)th}} moving boundary
may pass through \(Q_{\jmath q}\). Declare \(q=0\) bad in every column so that the undefined endpoint
\(a_j(\cdot,0)\) is not used, and set
\begin{shrinkeq}{0em}
\[
  \mathcal B_\jmath 
  :=
  \{0\}\cup\bigcup_{r=1}^R\mathcal B_{\jmath ,r},
  \qquad
  \mathcal B_k
  :=
  \left\{
    (\jmath ,q):
    0\le \jmath \le k-1,\ q\in\mathcal B_\jmath 
  \right\}.
\]
\end{shrinkeq}
Since \(K_q^{(k)}\) has length \(1/k\),
\(\lvert\mathcal B_{\jmath ,r}\rvert\le
k(\overline\zeta_{\jmath ,r}-\underline\zeta_{\jmath ,r})+2\). Hence
\begin{shrinkeq}{0em}
\begin{equation}\label{eq:bad-one-boundary}
  \lvert\mathcal B_k\rvert\le \sum_{\jmath =0}^{k-1}(1+\sum_{r=1}^R
  \lvert\mathcal B_{\jmath ,r}\rvert)
  \le
  k+\sum_{r=1}^R
  [
    k\sum_{\jmath =0}^{k-1}
    (\overline\zeta_{\jmath ,r}-\underline\zeta_{\jmath ,r})
    +2k
  ]
  \le(3R+1)k.
\end{equation}
\end{shrinkeq}
Here we used
\begin{shrinkeq}{0em}
\[
  \sum_{\jmath =0}^{k-1}
  (\overline\zeta_{\jmath ,r}-\underline\zeta_{\jmath ,r})
  \le
  \sum_{\jmath =0}^{k-1}V_{J_\jmath ^{(k)}}(\zeta_{r,N})
  \le
  V_{[0,1)}(\zeta_{r,N})
  \le
  V_{\mathbb R}(\eta_r)
  =
  1.
\]
\end{shrinkeq}
Since \(0\le a_{N,j}\le1\),
\begin{shrinkeq}{0em}
\begin{equation}\label{eq:bad-osc-gaussian}
  \sum_{(\jmath ,q)\in\mathcal B_k}\omega_{\jmath q}
  =
  \sum_{(\jmath ,q)\in\mathcal B_k}
  \operatorname{osc}_{Q_{\jmath q}}(a_{N,j})
  \le(3R+1)k.
\end{equation}
\end{shrinkeq}

\textbf{Bounding good cells.}
Fix \(\jmath \). If \(q\notin\mathcal B_\jmath \), then \(K_q^{(k)}\) misses
\([\underline\zeta_{\jmath ,r},\overline\zeta_{\jmath ,r}]\) for every \(r\); hence either
\(
  u<\zeta_{r,N}(v) \) for all \((v,u)\in Q_{\jmath q},
  \)
  or
  \(
  u>\zeta_{r,N}(v)\) for all \((v,u)\in Q_{\jmath q}.
\)
Since \(\zeta_{1,N}(v)\le\cdots\le\zeta_{R,N}(v)\), there is a unique
\(\ell\in\{0,\ldots,R\}\) such that for all \((v,u)\in Q_{\jmath q}\),
 $\zeta_{\ell,N}(v)<u<\zeta_{\ell+1,N}(v)$. 
By \eqref{eq:moving-strip-gaussian}, \(a_{N,j}(v,\cdot)\) therefore
has one common monotonicity direction throughout that cell. For fixed \(\jmath \), each \(\mathcal B_{\jmath ,r}\) is a contiguous collection
of row indices. Hence the complement of the bad rows induced by the boundary
has at most \(R+1\) contiguous blocks. Removing \(q=0\) cannot increase
that number. Denote the resulting good blocks by \(\mathcal Q_{\jmath ,1},\ldots,\mathcal Q_{\jmath ,M_\jmath }\) with
\(M_\jmath \le R+1\). On each block, \(u\mapsto a_{N,j}(v,u)\) has one
common monotonicity direction for every \(v\in J_\jmath ^{(k)}\). Consider first a good block on which
\(u\mapsto a_{N,j}(v,u)\) is nondecreasing.
For
\(q\in\mathcal Q_{\jmath ,\nu}\), we have \(q\ne0\), so
\(
  t_q:=q/k\in(0,1).
\)
Define $g_q(v):=a_{N,j}(v,t_q)$ and $U_{\jmath ,\nu}(u)
  :=
  \sup_{v\in J_\jmath ^{(k)}}a_{N,j}(v,u).$
Clearly, \(U_{\jmath ,\nu}\) is also nondecreasing on the block. Moreover, since \(t_q\in K_q^{(k)}\), we have
\(
\inf_{(v,u)\in Q_{\jmath q}}a_{N,j}(v,u)
=\inf_{v\in J_\jmath ^{(k)}}g_q(v)
\) and \(
U_{\jmath ,\nu}(t_q)=\sup_{v\in J_\jmath ^{(k)}}g_q(v).
\)
Therefore, adding and subtracting \(U_{\jmath ,\nu}(t_q)\) gives
\begin{shrinkeq}{0em}
\begin{equation}\label{eq:good-gaussian-cell}
\begin{aligned}
\omega_{\jmath q}
&= \sup_{(v,u)\in Q_{\jmath q}}a_{N,j}(v,u) - \inf_{(v,u)\in Q_{\jmath q}}a_{N,j}(v,u) = \sup_{u\in K_q^{(k)}}U_{\jmath ,\nu}(u)
  -\inf_{v\in J_\jmath ^{(k)}}g_q(v) \\
  &=\sup_{u\in K_q^{(k)}}U_{\jmath ,\nu}(u)-U_{\jmath ,\nu}(t_q)
 +\sup_{v\in J_\jmath ^{(k)}}g_q(v)-\inf_{v\in J_\jmath ^{(k)}}g_q(v) \le V_{K_q^{(k)}}(U_{\jmath ,\nu})+V_{J_\jmath ^{(k)}}(g_q).
\end{aligned}
\end{equation}
\end{shrinkeq}
For a nonincreasing good block retain the same left endpoint $t_q=q/k$
and the same $g_q$, but use the lower envelope 
\(
 D_{\jmath ,\nu}(u):=\inf_{v\in J_\jmath ^{(k)}}a_{N,j}(v,u).
\)
It is nonincreasing. Since the largest value in the row occurs at its
left endpoint, adding and subtracting $D_{\jmath ,\nu}(t_q)$ gives
\[
\begin{aligned}
 \omega_{\jmath q}
 &=\sup_{v\in J_\jmath ^{(k)}}g_q(v)-\inf_{u\in K_q^{(k)}}D_{\jmath ,\nu}(u)=\operatorname{osc}_{J_\jmath ^{(k)}}(g_q)
   +D_{\jmath ,\nu}(t_q)-\inf_{u\in K_q^{(k)}}D_{\jmath ,\nu}(u)\\
 &\le V_{J_\jmath ^{(k)}}(g_q)+V_{K_q^{(k)}}(D_{\jmath ,\nu}).
\end{aligned}
\]
Define $E_{\jmath ,\nu}=U_{\jmath ,\nu}$ on nondecreasing blocks and
$E_{\jmath ,\nu}=D_{\jmath ,\nu}$ on nonincreasing blocks (either choice on a constant
block). Both orientations therefore satisfy
\begin{equation}\label{eq:good-cell-both}
 \omega_{\jmath q}\le V_{J_\jmath ^{(k)}}(g_q)+V_{K_q^{(k)}}(E_{\jmath ,\nu}).
\end{equation}
The row sample $t_q$ is independent of the column and its orientation.
No endpoint value at $u=0$ or $u=1$ is used; for the last half-open row,
variation is understood as the supremum over finite partitions. For the contribution in the state direction, we have
\begin{shrinkeq}{0em}
\begin{equation}\label{eq:state-contribution-gaussian}
\begin{aligned}
\sum_{\substack{(\jmath ,q)\notin\mathcal B_k}}
V_{J_\jmath ^{(k)}}(g_q)
\le \sum_{q=1}^{k-1}\sum_{\jmath =0}^{k-1} V_{J_\jmath ^{(k)}}
  \bigl(a_j(Z_{N,j-1}(\cdot),t_q)\bigr)
\le \sum_{q=1}^{k-1}V_{\mathbb R}
  \bigl(a_j(\cdot,t_q)\bigr)
\le (k-1)B_{\pi}.
\end{aligned}
\end{equation}
\end{shrinkeq}
Here \(\xi=\sigma_jH^{-1}(u)\) ranges over \(\mathbb R\) as
\(u\in(0,1)\), so \eqref{eq:bpi} gives
\(V_{\mathbb R}(a_j(\cdot,u))\le B_\pi\). On each good block, use the
orientation-dependent envelope $E_{\jmath ,\nu}$ defined above. It is monotone
and takes values in $[0,1]$. Define
\(I_{\jmath ,\nu}:=\bigcup_{q\in\mathcal Q_{\jmath ,\nu}}K_q^{(k)}\). Additivity of total variation over adjacent intervals gives
\begin{shrinkeq}{0em}
\begin{equation}\label{eq:u-contribution-gaussian}
\begin{aligned}
\sum_{\jmath =0}^{k-1}\sum_{\nu=1}^{M_\jmath }
\sum_{q\in\mathcal Q_{\jmath ,\nu}}V_{K_q^{(k)}}(E_{\jmath ,\nu})
&\le \sum_{\jmath =0}^{k-1}\sum_{\nu=1}^{M_\jmath }
V_{I_{\jmath ,\nu}}(E_{\jmath ,\nu})\le \sum_{\jmath =0}^{k-1}M_\jmath \le k(R+1).
\end{aligned}
\end{equation}
\end{shrinkeq}
Combining
\eqref{eq:bad-osc-gaussian}, \eqref{eq:good-cell-both},
\eqref{eq:state-contribution-gaussian}, and
\eqref{eq:u-contribution-gaussian}, we obtain
\begin{shrinkeq}{0em}
\[
\begin{aligned}
  \sum_{\jmath ,q=0}^{k-1}\omega_{\jmath q}
  \le
  (3R+1)k
  +
  kB_\pi
  +
  k(R+1)
  =
  k(B_\pi+4R+2).
\end{aligned}
\]
\end{shrinkeq}
Dividing by \(k\) and taking the supremum over \(k\ge1\) gives $C_{a,j} \le B_\pi+4R+2$,
which is the uniform grid-oscillation estimate used in the boundary-tube argument.

Write $A_j:=\int_0^1V_{\mathbb R}(a_j(\cdot,u))\,\dd u$. Continuity of $\pi$ and $H^{-1}$ makes $a_j$ continuous on
$\mathbb R\times(0,1)$; its state variation is the supremum over countably
many rational finite partitions and is therefore measurable in $u$.
Finally, \(A_j=\int_0^1V_{\mathbb R}(a_j(\cdot,u))\,\mathrm{d}u\le B_{\pi}\).

For the proposal in \eqref{eq:symmetric-proposal},
\begin{equation}\label{eq:proposal-threshold-variation}
 \beta_{j,x}(z)=H((x-z)/\sigma_j),\qquad
 C_{\beta,j}=\sup_{x\in\mathbb R}V_{\mathbb R}(\beta_{j,x})=1.
\end{equation}
For every fixed $x$, the displayed function is nonincreasing in $z$,
with limits $1$ and $0$ as $z$ tends to $-\infty$ and $+\infty$.
Its variation is therefore exactly one. Since $Z_{N,j-1}$ is
nondecreasing, $V_{[0,1)}(\beta_{j,x}\circ Z_{N,j-1})\le1$ for every
sorted array, independently of $N$ and $x$.

\Cref{prop:kernel-variation} gives $\Lambda_j\le B_\pi+2$.
Hence all three quantitative conditions of \Cref{thm:main} hold.
Together with its point-set and initialization conditions, this proves
the stated fixed-horizon rate.
\end{proof}

\section{Verification of target classes}
\label{app:targets}
We verify the hypotheses of \Cref{prop:symmetric-location} for all target
classes summarized in \Cref{tab:state-variation} of the main paper.

\subsection{Positive log-concave targets}
\begin{lemma}[Positive log-concave targets]
\label{lem:logconcave}
A finite-valued, strictly positive log-concave probability density on
$\mathbb R$ satisfies the target hypotheses of \Cref{prop:symmetric-location} with $R=1$ and
$B_\pi\le1$. Consequently, under \Cref{ass:points}, the BV sampling-error conclusion of \Cref{thm:main} holds for every proposal
in \eqref{eq:symmetric-proposal}, without target differentiability.
\end{lemma}
\begin{proof}
The finite concave function $\ell=\log\pi$ is continuous and locally
absolutely continuous. Its derivative exists almost everywhere and is
nonincreasing. For $\xi\in\mathbb R$, set
$\alpha_\xi(x):=1\wedge\exp\{\ell(x+\xi)-\ell(x)\}$. For $\xi>0$,
$\ell(x+\xi)-\ell(x)=\int_0^\xi\ell'(x+t)\,\dd t$ is nonincreasing in $x$;
for negative $\xi$ the direction reverses. Composing with $u\mapsto1\wedge e^u$
gives $V_{\mathbb R}(\alpha_\xi)\le1$, and thus $B_\pi\le1$.
Integrability supplies a negative secant slope somewhere to the right
and a positive one somewhere to the left: otherwise the positive density
would be bounded below on a half-line. Concavity then bounds both tails
by decreasing exponentials towards their respective infinities.
The continuous density thus attains its maximum on a compact interval.
Concavity makes it nondecreasing before and nonincreasing after any selected
point of that mode interval. This provides one finite partition breakpoint,
including when the mode is flat.
\end{proof}
This includes the standard Gaussian and the nondifferentiable Laplace
targets. Substituting $R=1$ and $B_\pi\le1$ into
\Cref{prop:symmetric-location} gives $C_{a,j}\le7$, $C_{\beta,j}=1$
and $\Lambda_j\le3$, as recorded in \Cref{tab:state-variation}.

\subsection{Student-\texorpdfstring{$t$}{t} targets}
\begin{lemma}[Exact Student-$t$ acceptance variation]
\label{lem:student-variation}
For the Student-$t_\nu$ target, $\nu>0$, the quantity in \eqref{eq:bpi}
is $B_\pi=2$. Hence one may take
$C_{a,j}\le8$, $C_{\beta,j}=1$, $\Lambda_j\le4$.
\end{lemma}
\begin{proof}
Write $a=(\nu+1)/2$ and fix $\xi>0$. The uncapped ratio is
$r_\xi(x)=\{(\nu+x^2)/(\nu+(x+\xi)^2)\}^a$, which is at least one exactly
when $x\le-\xi/2$. Its logarithmic derivative is
\[
 \frac{\mathrm d}{\mathrm dx}\log r_\xi(x)
 =\frac{(\nu+1)\xi(x^2+\xi x-\nu)}{(\nu+x^2)(\nu+(x+\xi)^2)}.
\]
The numerator has two distinct real roots since $\xi^2+4\nu>0$.
Of these, the negative root lies in the
capped region; the other is $x_*=(R_\xi-\xi)/2$ with $R_\xi=\sqrt{\xi^2+4\nu}$.
Thus $\alpha_\xi=1\wedge r_\xi$ decreases from one after $-\xi/2$ to a single
minimum at $x_*$ and then increases to one as $x\to\infty$.
Since $x_*(x_*+\xi)=\nu$,
\[
 \alpha_\xi(x_*)=\left(\frac{R_\xi-\xi}{R_\xi+\xi}\right)^a,
 \qquad V_{\mathbb R}(\alpha_\xi)
 =2\left[1-\left(\frac{R_\xi-\xi}{R_\xi+\xi}\right)^a\right].
\]
Evenness of the density gives $\alpha_{-\xi}(x)=\alpha_\xi(-x)$, and at
$\xi=0$ the variation is zero. The displayed values tend to $2$ as
$\xi\to\infty$, proving the supremum. The density is increasing on
$(-\infty,0)$ and decreasing on $(0,\infty)$, so one may take $R=1$.
Substitution of $B_\pi=2$, $R=1$ into the bounds in
\Cref{prop:symmetric-location} gives the constants.
\end{proof}

\subsection{Score-root criterion and the double-well target}
\begin{lemma}[Score-root criterion]
\label{lem:state-variation-roots}
Let \(\pi:\mathbb R\to(0,\infty)\) be a continuously differentiable density,
let \(\ell:=\log\pi\), and set
\(S_\xi(x):=\ell'(x+\xi)-\ell'(x)\). If some integer \(K\ge0\) bounds the number of
distinct real zeros of \(S_\xi\) for every \(\xi\ne0\), then
\begin{shrinkeq}{0em}
\[
B_\pi\le K+1<\infty.
\]
\end{shrinkeq}
\end{lemma}
\begin{proof}
For $\xi\in\mathbb R$, set $\alpha_\xi(x):=1\wedge (\pi(x+\xi)/\pi(x))
=1\wedge\exp\{\ell(x+\xi)-\ell(x)\}.$
If $\xi=0$, then $\alpha_0\equiv1$ and $V_{\mathbb R}(\alpha_0)=0$. Fix $\xi\ne0$,
let $K_\xi\le K$ be the actual number of distinct real zeros of $S_\xi$, and
write them as $\tau_{\xi,1}<\cdots<\tau_{\xi,K_\xi}$. By continuity, $S_\xi$ has a
fixed sign on each component of
$\mathbb R\setminus\{\tau_{\xi,1},\ldots,\tau_{\xi,K_\xi}\}$.
Hence $x\mapsto\ell(x+\xi)-\ell(x)$, and therefore $\alpha_\xi$,
is monotone on each of the resulting $K_\xi+1$ intervals.
Since $0<\alpha_\xi\le1$, its variation on each such interval is at most one.
The additivity of total variation gives
\(V_{\mathbb R}(\alpha_\xi)\le K_\xi+1\le K+1\).
Taking the supremum over \(\xi\ne0\) proves the result.
\end{proof}

Under the remaining hypotheses of \Cref{prop:symmetric-location},
including a finite monotonicity partition with $R$ breakpoints and the
specified symmetric-location proposal, \Cref{lem:state-variation-roots}
gives $B_\pi\le K+1$. Substituting this bound into
\Cref{prop:symmetric-location} yields
\[
 C_{a,j}\le K+4R+3,\quad C_{\beta,j}=1,
 \quad \Lambda_j\le K+3.
\]
For the double-well target, direct differentiation gives, for $\xi\ne0$,
\begin{equation}\label{eq:score-examples}
 S_\xi^{\rm DW}(x)=-4\xi(3x^2+3\xi x+\xi^2-1).
\end{equation}
This polynomial has at most two distinct real roots, so
\Cref{lem:state-variation-roots} gives $B_\pi\le3$ with $K=2$.
The density has critical points $-1,0,1$, yielding a monotonicity
partition with $R=3$. Consequently, \Cref{prop:symmetric-location} gives
$C_{a,j}\le17$, $C_{\beta,j}=1$ and $\Lambda_j\le5$,
which verifies the double-well row of \Cref{tab:state-variation}.

\subsection{Finite Gaussian mixtures}
\label{sec:supp-gaussian-mixture}

Fix a positive integer $J$ for the number of components, independently of
$N=b^m$. We use the unnormalized family
\[
 \Theta_J=\{(\bm{w},\bm{a},\bm{\mu})\in\mathbb R^{3J}:w_r>0,\ a_r>0\ (1\le r\le J)\},
 \qquad p_{\bm{\theta}}(x)=\sum_{r=1}^J w_r e^{-a_r(x-\mu_r)^2}.
\]
All parameters are finite real numbers. For every $\bm{\theta}\in\Theta_J$,
$Z_{\bm{\theta}}=\int_{\mathbb R}p_{\bm{\theta}}$ is positive and finite and
$\pi_{\bm{\theta}}=p_{\bm{\theta}}/Z_{\bm{\theta}}$ is a probability density.

\begin{proposition}[Parameter-uniform Gaussian mixture bounds]
\label{prop:supp-gaussian-mixture}
For each fixed $J\ge1$ there exist finite integers $R_J,K_J$, depending
only on $J$, such that, simultaneously for every $\bm{\theta}\in\Theta_J$,
\[
 \#\{x:p'_{\bm{\theta}}(x)=0\}\le R_J,\qquad
 \#\{x:\ell'_{\bm{\theta}}(x+\xi)-\ell'_{\bm{\theta}}(x)=0\}\le K_J\quad(\xi\ne0),
 \qquad \ell_{\bm{\theta}}=\log\pi_{\bm{\theta}}.
\]
Consequently, for the symmetric location proposals of
\eqref{eq:symmetric-proposal}, uniformly in $\bm{\theta}$ and the deterministic
positive scales,
\begin{equation}\label{eq:mixture-uniform-constants}
\begin{aligned}
 B_{\pi_{\bm{\theta}}},A_j&\le K_J+1,& C_{a,j}&\le K_J+4R_J+3,\\
 C_{\beta,j}&=1,& \Lambda_j&\le K_J+3,&
 M_j&\le D_J:=14K_J+56R_J+132.
\end{aligned}
\end{equation}
With \Cref{ass:points} at $p=1$, $D=3$, these yield, for fixed $T$, a constant
$C_{J,b,t,T}<\infty$ and a cutoff depending only on $b,t$ such that
\[
 \mathbb E_{\mathrm{RQMC}}[(\widehat\mu_{N,T}(f)-\mu_T(f))^2]
 \le C_{J,b,t,T}V_{\mathbb R}(f)^2N^{-4/3}
\]
for all pointwise BV $f$, uniformly also in the deterministic initial state
$x_0$ and $0\le T_0<T$. The integers $R_J,K_J$ are nonquantitative existence
bounds; no compact restriction on the parameter domain is imposed.
\end{proposition}
\begin{proof}
We use three precise standard facts. The real exponential field
$\mathbb R_{\exp}$ is o-minimal, so every definable subset of the line is
a finite union of points and intervals
\citep[Definition~1.2 and Example~1.6]{loi2010ominimal}. Its definable sets
are closed under the finite operations used below
\citep[Proposition~2.2 and Theorem~2.3]{loi2010ominimal}.
Finally, if $A\subset\mathbb R^{n+1}$ is definable and each fiber
$A_{\bm{y}}=\{x:(\bm{y},x)\in A\}$ is finite, then a single integer bounds
$\#A_{\bm{y}}$ for all $\bm{y}\in\mathbb R^n$
\citep[Theorem~2.10 (Un), p.~24]{loi2010ominimal}.

Both $p_{\bm{\theta}}(x)$ and
\[
 p'_{\bm{\theta}}(x)=-2\sum_{r=1}^J w_ra_r(x-\mu_r)e^{-a_r(x-\mu_r)^2}
\]
are definable jointly in $(\bm{\theta},x)$ and analytic in $x$. We also use the
real-analytic identity theorem: an analytic function on a connected open
interval that vanishes on a nonempty open subinterval vanishes identically.
For completeness, its local power series shows that a zero of a
nonidentically-zero analytic germ is isolated (factor out the first nonzero
power). Thus the set of points with a zero germ is both open and closed:
a limit of such points cannot have a nonzero germ, whose zeros would be
isolated. Connectedness proves the stated identity theorem.

Consider the single definable set
\[
 \mathcal T_J=\{(\bm{\theta},x):\bm{\theta}\in\Theta_J,\ p'_{\bm{\theta}}(x)=0\}.
\]
If an admissible fiber were infinite, o-minimality would give an interval
of zeros and the identity theorem would give $p'_{\bm{\theta}}\equiv0$.
Then $p_{\bm{\theta}}$ would be a positive constant on the line, contradicting
integrability. All admissible fibers are therefore finite; all inadmissible
fibers are empty. Uniform finiteness gives $R_J$. Between successive
critical points, the continuous derivative has a fixed sign, giving the
required finite monotonicity partition.

For the score difference define
\[
\begin{aligned}
 H_{\bm{\theta}}(\xi,x)&=p'_{\bm{\theta}}(x+\xi)p_{\bm{\theta}}(x)-p_{\bm{\theta}}(x+\xi)p'_{\bm{\theta}}(x),\\
 \mathcal Z_J&=\{(\bm{\theta},\xi,x):\bm{\theta}\in\Theta_J,\ \xi\ne0,\ H_{\bm{\theta}}(\xi,x)=0\}.
\end{aligned}
\]
This set is definable jointly in all its variables. Positivity of $p_{\bm{\theta}}$
makes its roots exactly the score roots. By definition, fibers at $\xi=0$
or outside $\Theta_J$ are empty; the identically zero function
$H_{\bm{\theta}}(0,\cdot)$ is not a fiber of the root problem.
Fix an admissible $\bm{\theta}$ and $\xi\ne0$. If the fiber were infinite,
o-minimality and analyticity would imply $H_{\bm{\theta}}(\xi,\cdot)\equiv0$. Hence
\[
 \frac{\dd}{\dd x}\frac{p_{\bm{\theta}}(x+\xi)}{p_{\bm{\theta}}(x)}
 =\frac{H_{\bm{\theta}}(\xi,x)}{p_{\bm{\theta}}(x)^2}=0,
 \qquad p_{\bm{\theta}}(x+\xi)=c\,p_{\bm{\theta}}(x)
\]
for a positive constant $c$. Integration over the line gives $c=1$.
The positive continuous function $p_{\bm{\theta}}$ would then have period $|\xi|$,
so infinitely many disjoint periods, each with the same positive integral,
would contradict $Z_{\bm{\theta}}<\infty$. Thus every fiber over $(\bm{\theta},\xi)$
is finite. Only now apply uniform finiteness to $\mathcal Z_J$ to obtain
$K_J$ simultaneously in the parameters and all nonzero shifts.
Normalization cancels from scores and ratios, so no definability of the
normalizing integral is needed.

\Cref{lem:state-variation-roots,prop:symmetric-location} give the
bounds on $B_{\pi_{\bm{\theta}}}$, $C_{a,j}$, $C_{\beta,j}$ and $\Lambda_j$.
The state-variation bound in the symmetric-location proof gives
$V_{\mathbb R}(a_j(\cdot,u))\le B_{\pi_{\bm{\theta}}}$ for $0<u<1$, hence
$A_j=\int_0^1V_{\mathbb R}(a_j(\cdot,u))\,\dd u
\le B_{\pi_{\bm{\theta}}}\le K_J+1$.
Together with the local coefficient \eqref{eq:B-tube-bound}, this proves
\eqref{eq:mixture-uniform-constants}.
The common tube-radius range is $0<\varepsilon<1/2$, so
the proof of \Cref{thm:main} in \Cref{app:onedim} gives the common cutoff \eqref{eq:main-cutoff}.
Let {{$C_{T_0,T}(f)$}} denote the coefficient of $N^{-4/3}$ in
\eqref{eq:main-rate}. With $M_s\le D_J$, $\Lambda_q\le K_J+3$ and
$c_*:=c_{b,t,3}D_J$, its explicit error constant satisfies
\[
 {{C_{T_0,T}(f)}}
 \le\frac{c_*V_{\mathbb R}(f)^2}{L^2}
 \sum_{s=1}^T
 \left[\sum_{j=\max\{s,T_0+1\}}^T(K_J+3)^{j-s}\right]^2.
\]
Each inner sum has at most $L$ terms and each term is at most
$(K_J+3)^{T-s}$. Thus one may take
$C_{J,b,t,T}:=c_*\sum_{q=0}^{T-1}(K_J+3)^{2q}$ in
\Cref{prop:supp-gaussian-mixture}. This is uniform in $T_0$ and $x_0$,
as well as in all the mixture parameters and positive proposal scales.
\end{proof}
Coincident components and parameters arbitrarily close to the excluded
positivity boundary do not create infinite fibers within $\Theta_J$.
This proof supplies neither explicit root counts nor uniform mixing,
stationary bias, polynomial moments or bounds for unbounded component count.

The experimental target
\[
 \widetilde\pi(x)=e^{-2(x+1.5)^2}+0.8e^{-3x^2}+0.6e^{-4(x-1.5)^2}
\]
has $J=3$, $\bm{w}=(1,0.8,0.6)$, $\bm{a}=(2,3,4)$ and
$\bm{\mu}=(-1.5,0,1.5)$, all in the admissible parameter domain.
It is therefore covered by \Cref{prop:supp-gaussian-mixture}, with the
constants \eqref{eq:mixture-uniform-constants}; no numerical root counting
is needed for this analytical example.

\section{Product targets and coordinatewise MWG}
\label{sec:supp-product-mwg}
We prove \Cref{cor:product-mwg,cor:polynomial-tests}.
Throughout this section, $j$ indexes
complete sweeps, each of which updates coordinates $r=1,\ldots,p$
once in that deterministic order. All scalar kernels, proposal
parameters and underlying nets are deterministic; only the independent
randomizations of the nets are random. In particular, proposal scales
and update schedules do not adapt to other coordinates.

\subsection{Proof of \texorpdfstring{\Cref{cor:product-mwg}}{Corollary \getrefnumber{cor:product-mwg}}}
\label{subsec:supp-product-mwg-proof}
By \eqref{eq:mwg-acceptance}, the $r$th update is the scalar MH kernel
associated with $(\pi_r,q_{r,j})$. Starting from $\pi_r(x_{0,r})>0$, we need only verify
that the Array-RQMC marginal evolution is unaffected by other
coordinate updates.

Sorting full particle vectors only permutes the
values of coordinates that are not being updated.
Rank matching pairs the sorted old particles with
the sorted RQMC points bijectively. Exact
first-coordinate stratification additionally
ensures the rank-interval identity in the scalar
analogue of \eqref{eq:hilbert-step}, as required
by the one-dimensional analysis.

Consequently, before updating coordinate $r$
in sweep $j$, the $r$th marginal multiset is
unchanged from the preceding sweep:
\[
 \{X_{i,j,r}^{(r-1)}:0\le i<N\}
 =
 \{X_{i,j-1,r}:0\le i<N\},
\]
where equality is understood with multiplicities. The permutation $\varsigma_{r,j}$ orders these
values nondecreasingly:
\(
 X_{\varsigma_{r,j}(0),j,r}^{(r-1)}
 \le\cdots\le
 X_{\varsigma_{r,j}(N-1),j,r}^{(r-1)}.
\)
Later coordinate updates only permute the
resulting $r$th values. Hence, as multisets,
\begin{equation}\label{eq:product-multiset-recursion}
 \{X_{i,j,r}:0\le i<N\}
 =
 \left\{
 \Psi_{r,j}\!\left(
 X_{\varsigma_{r,j}(i),j,r}^{(r-1)},
 U_{(i),r,j},\Omega_{(i),r,j}
 \right):0\le i<N
 \right\}.
\end{equation}
Starting from $X_{i,0,r}=x_{0,r}$, this is exactly the one-dimensional
Array-RQMC recursion with natural sorting. Since the initial particles are deterministic,
\eqref{eq:product-multiset-recursion} implies by induction
that, for each coordinate $r$, the sequence of marginal
multisets $\{X_{i,j,r}:0\le i<N\}$, $j=0,\ldots,T$,
depends only on the point sets
$P_{N,r,1},\ldots,P_{N,r,T}$. In particular, with $L=T-T_0$, the marginal estimator
\[
 \widehat\mu_{r,N,T}(f_r)
 :=\frac1{NL}
 \sum_{j=T_0+1}^T\sum_{i=0}^{N-1}
 f_r(X_{i,j,r})
\]
depends only on these point sets.
Since the point-set randomizations are independent
across coordinates, the marginal estimators are
mutually independent, even though the coordinates
of a labelled full particle need not be independent.

Let $\nu_{r,j}$ be the law of the scalar IID-driven MH chain after
$j$ updates from $x_{0,r}$, with kernels corresponding to
$q_{r,1},\ldots,q_{r,j}$. It is also the $r$th marginal law of ordinary
systematic-scan MWG at the end of sweep $j$: other coordinate moves do
not change that coordinate, and its transition depends only on its
own old value. Put
$\mu_{r,T}(f_r)=L^{-1}\sum_{j=T_0+1}^T\nu_{r,j}(f_r)$.
The scalar unbiasedness identity \eqref{eq:one-time-unbias} gives
$\mathbb E_{\mathrm{RQMC}}\widehat\mu_{r,N,T}(f_r)=\mu_{r,T}(f_r)$.
For additive $f$, the sweep-endpoint estimator and its target satisfy
\[
 \widehat\mu_{N,T}(f)=\sum_{r=1}^p \widehat\mu_{r,N,T}(f_r),
 \qquad \mu_T(f)=\sum_{r=1}^p \mu_{r,T}(f_r).
\]
Thus the marginal errors
$e_r=\widehat\mu_{r,N,T}(f_r)-\mu_{r,T}(f_r)$ are independent and
centered. For pointwise BV tests they are square integrable, so
\begin{equation}\label{eq:product-error-identity}
 \mathbb E_{\mathrm{RQMC}}
 [(\widehat\mu_{N,T}(f)-\mu_T(f))^2]
 =\sum_{r=1}^p \mathbb E_{\mathrm{RQMC}}[e_r^2].
\end{equation}
The identity holds for every admissible $N$.

Apply \Cref{thm:main} separately to each scalar recursion. To make the
constant explicit, write $C_{a,r,s},C_{\beta,r,s},\Lambda_{r,s}$ for its
three quantitative coefficients, and
$\mathcal H_{r,s}^{T_0,T}$ for the propagation weight
\eqref{eq:horizon-weight} computed from $\Lambda_{r,s}$. Then
\eqref{eq:innovation-variance} gives
\begin{equation}\label{eq:product-coordinate-constant}
 {{C_{r,T_0,T}(f_r)}}=\frac{c_{b,t,3}V_{\mathcal X_r}(f_r)^2}{L^2}
 \sum_{s=1}^T(14C_{a,r,s}+6C_{\beta,r,s}+84)
                   (\mathcal H_{r,s}^{T_0,T})^2.
\end{equation}
The cutoff \eqref{eq:main-cutoff} depends only on the common $b,t$,
since every scalar tube bound uses the same radius range
$0<\varepsilon<1/2$. For every admissible $N\ge N_0$,
$\mathbb E_{\mathrm{RQMC}}[e_r^2]\le {{C_{r,T_0,T}(f_r)}}N^{-4/3}$ simultaneously
for all coordinates. Substitution in \eqref{eq:product-error-identity}
proves \eqref{eq:product-mwg-rate}. For a family with $b,t,T,T_0$ fixed
independently of $p$, a dimension-uniform bound follows whenever
$\sup_{p\ge1}\sum_{r=1}^p {{C_{r,T_0,T}(f_r)}}<\infty$, with the same cutoff
$N_0$. 

\subsection{Proof of \texorpdfstring{\Cref{cor:polynomial-tests}}{Corollary \getrefnumber{cor:polynomial-tests}}}
\label{sec:supp-polynomial}
We first extend the time-average BV bound in \Cref{thm:main} directly by
$L^2$ approximation. Write $L=T-T_0$ and define the average target measure
$\bar\nu_{T_0,T}:=L^{-1}\sum_{j=T_0+1}^T\nu_j$, so that
$\mu_T(f)=\bar\nu_{T_0,T}(f)$. Set
\[
 A_{T,T_0}^2:=\frac{c_{b,t,3}}{L^2}
 \sum_{s=1}^T M_s(\mathcal H_s^{T_0,T})^2,
 \qquad M_s=14C_{a,s}+6C_{\beta,s}+84.
\]
For a random variable $Y$, write
$\|Y\|_{2,\mathrm{RQMC}}=(\mathbb E_{\mathrm{RQMC}}[|Y|^2])^{1/2}$.

\begin{proposition}[Time-average BV approximation inequality]
\label{prop:bv-approximation}
Under the sampling and kernel assumptions of \Cref{thm:main}, let
$f\in L^2(\bar\nu_{T_0,T})$. For every admissible $N\ge N_0$,
with $N_0$ from \eqref{eq:main-cutoff} in \Cref{app:onedim},
\begin{equation}\label{eq:bv-approx-step}
\begin{aligned}
 \|\widehat\mu_{N,T}(f)-\mu_T(f)\|_{2,\mathrm{RQMC}}\le\inf_{g\in\mathrm{BV}(\mathcal X)\cap L^2(\bar\nu_{T_0,T})}
 \left\{A_{T,T_0}N^{-2/3}V_{\mathcal X}(g)
       +\|f-g\|_{L^2(\bar\nu_{T_0,T})}\right\}.
\end{aligned}
\end{equation}
\end{proposition}
\begin{proof}
For an admissible $g$, put $h=f-g$. The estimator is a positive,
normalized empirical average and is unbiased for $\bar\nu_{T_0,T}$ by
\eqref{eq:one-time-unbias}. Jensen's inequality therefore gives
\[
\begin{aligned}
 \|\widehat\mu_{N,T}(h)-\mu_T(h)\|_{2,\mathrm{RQMC}}^2
 &=\operatorname{Var}_{\mathrm{RQMC}}(\widehat\mu_{N,T}(h))\\
 &\le\mathbb E_{\mathrm{RQMC}}[\widehat\mu_{N,T}(h)^2]
 \le\mathbb E_{\mathrm{RQMC}}[\widehat\mu_{N,T}(h^2)]
 =\bar\nu_{T_0,T}(h^2).
\end{aligned}
\]
Apply \Cref{thm:main} to $g$, use the triangle inequality in $L^2$,
and take the infimum over $g$ to obtain \eqref{eq:bv-approx-step}.
\end{proof}

\begin{proposition}[Scalar polynomial extension]
\label{prop:scalar-polynomial}
Under the sampling and kernel assumptions of \Cref{thm:main} on $\mathcal X=\mathbb R$,
suppose $Y_j(z,u)=z+\sigma\Phi^{-1}(u)$, where $0<u<1$,
$\sigma>0$ is fixed and $\Phi$ is the standard normal CDF.
For fixed integers $0\le T_0<T$ and $\ell\ge1$, let $f_\ell(x)=x^\ell$.
Then, for sufficiently large admissible $N$,
\begin{equation}\label{eq:polynomial-test-rate}
 \mathbb E_{\mathrm{RQMC}}
 [(\widehat\mu_{N,T}(f_\ell)-\mu_T(f_\ell))^2]
 =\mathcal O_{T,\ell}(N^{-4/3}(\log N)^\ell).
\end{equation}
\end{proposition}

\begin{proof}
We write $\mathbb E_{\mathrm{MC}}$ for expectation under the ordinary
IID-driven chain.
Let \(\xi_k\in\{0,1\}\) be the corresponding acceptance indicators.  Then
\[
  X_j=x_0+\sigma\sum_{k=1}^{j}\xi_k\mathcal Z_k
  \quad\Longrightarrow\quad
  |X_j|\leq|x_0|+\sigma\sum_{k=1}^{T}|\mathcal Z_k|,
  \qquad j\leq T,
\]
where
$U_k\stackrel{\mathrm{iid}}{\sim}\Unif(0,1)$ are the proposal uniforms of
the ordinary chain and
$\mathcal Z_k=\Phi^{-1}(U_k)\stackrel{\mathrm{iid}}{\sim}\mathcal N(0,1)$.
Consequently, for every fixed \(T\), there exists
\(0<\kappa_T<1/(4\sigma^2T)\) such that
\[
\begin{aligned}
 \E_{\mathrm{MC}}[e^{\kappa_TX_j^2}]
 \leq e^{2\kappa_Tx_0^2}
 \prod_{k=1}^{T}\E[e^{2\kappa_T\sigma^2T\mathcal Z_k^2}]\leq e^{2\kappa_Tx_0^2}
 (1-4\kappa_T\sigma^2T)^{-T/2}
 =M_T<\infty.
\end{aligned}
\]
Here we used
\(\E[e^{c\mathcal Z^2}]=(1-2c)^{-1/2}\) for
\(\mathcal Z\sim\mathcal N(0,1)\) and \(c<1/2\). For \(R>0\), define
\[
  f_{\ell ,R}(x):=(\max\{-R,\min\{x,R\}\})^\ell .
\]
Then \(f_{\ell ,R}\) has finite total variation
\(V_{\mathbb R}(f_{\ell ,R})\leq2R^\ell \), and
\[
\begin{aligned}
 |f_\ell (x)-f_{\ell ,R}(x)|^2
 &\leq(|x|^\ell -R^\ell )_+^2
 \leq|x|^{2\ell }\one_{\{|x|>R\}}\\
 &=\left(|x|^{2\ell }e^{-\kappa_Tx^2/2}\right)
   \left(e^{-\kappa_Tx^2/2}\one_{\{|x|>R\}}\right)e^{\kappa_Tx^2}\\
 &\leq\left(\frac{2\ell }{e\kappa_T}\right)^\ell 
 e^{-\kappa_TR^2/2}e^{\kappa_Tx^2}.
\end{aligned}
\]
Thus
\[
\begin{aligned}
 \|f_\ell -f_{\ell ,R}\|_{L^2(\nu_j)}^2
 &=\E_{\mathrm{MC}}[|f_\ell (X_j)-f_{\ell ,R}(X_j)|^2]\leq\left(\frac{2\ell }{e\kappa_T}\right)^\ell 
 e^{-\kappa_TR^2/2}\E_{\mathrm{MC}}[e^{\kappa_TX_j^2}]\\
 &\leq\left(\frac{2\ell }{e\kappa_T}\right)^\ell M_T
 e^{-\kappa_TR^2/2}
 =C_{T,\ell }e^{-\kappa_TR^2/2}.
\end{aligned}
\]

Averaging this bound over $j=T_0+1,\ldots,T$ yields
\[
 \|f_\ell-f_{\ell,R}\|_{L^2(\bar\nu_{T_0,T})}
 \le C_{T,\ell}^{1/2}e^{-\kappa_TR^2/4}.
\]
For admissible $N\ge\max(N_0,2)$, choose
$R^2=8\log N/(3\kappa_T)$. Applying \Cref{prop:bv-approximation}
directly to the time-average estimator gives
\[
\begin{aligned}
 \|\widehat\mu_{N,T}(f_\ell)-\mu_T(f_\ell)\|_{2,\mathrm{RQMC}}
 &\le 2A_{T,T_0}R^\ell N^{-2/3}
       +C_{T,\ell}^{1/2}e^{-\kappa_TR^2/4}\\
 &=\mathcal O_{T,\ell}\!\left(N^{-2/3}(\log N)^{\ell/2}\right).
\end{aligned}
\]
Squaring proves \eqref{eq:polynomial-test-rate}.
The constants depend on the fixed horizon, proposal scale and initial
state. The argument uses Gaussian proposal tails and controls sampling
error relative to the finite-horizon target.
\end{proof}

We now prove \Cref{cor:polynomial-tests}. Suppose $\mathcal X_r=\mathbb R$,
each coordinatewise one-dimensional MH construction satisfies the
hypotheses of \Cref{prop:scalar-polynomial}, and its proposal is
$Y_{r,j}(x,u)=x+\sigma_r\Phi^{-1}(u)$ with fixed $\sigma_r>0$.
The finite-horizon Gaussian-increment estimate in
\Cref{sec:supp-polynomial} gives finite moments of every order under
$\nu_{r,j}$. Scalar unbiasedness, first for nonnegative truncations and
then by monotone convergence {{(see \Cref{app:innovation-identities} of the main paper)}}, applies to each integrable polynomial
and to its square. Jensen's inequality for the empirical average then
shows that its estimator has a finite second moment. Consequently the
centering and independence argument above remains valid for
$f_r(x)=x^{\ell_r}$ with any fixed integer $\ell_r\ge1$.
For $f(\bm{x})=\sum_r c_rf_r(x_r)$, linearity and independence give
\begin{equation}\label{eq:product-weighted-error-identity}
 \mathbb E_{\mathrm{RQMC}}[(\widehat\mu_{N,T}(f)-\mu_T(f))^2]
 =\sum_{r=1}^p c_r^2\mathbb E_{\mathrm{RQMC}}[e_r^2].
\end{equation}
By \Cref{prop:scalar-polynomial}, for fixed $p$ there are finite constants
${{K_{r,T_0,T,\ell_r}}}$ such that, for sufficiently large admissible $N$,
\begin{equation}\label{eq:product-polynomial-rate}
 \mathbb E_{\mathrm{RQMC}}
 [(\widehat\mu_{N,T}(f)-\mu_T(f))^2]
 \le N^{-4/3}\sum_{r=1}^p c_r^2{{K_{r,T_0,T,\ell_r}}}(\log N)^{\ell_r}.
\end{equation}
Finite linear combinations of powers within each coordinate follow
by the triangle inequality in $L^2$ before combining coordinate errors.
Equation~\eqref{eq:product-polynomial-rate} proves
\eqref{eq:additive-polynomial-rate} and \Cref{cor:polynomial-tests};
when all $\ell_r=\ell$, the finite sum gives the stated common-degree
rate. This extends the additive result to unbounded polynomial tests
without treating them as BV. As in the scalar bound, the constants can depend
on the initial states, proposal scales and fixed horizon.

\section{Comparison with one-dimensional inversion}
\label{app:inversion}

The same local-indicator framework also applies
to one-dimensional inverse-transform transitions {{\citep[Proposition~6]{lecuyer2008}}},
with a two-dimensional Array-RQMC point set
rather than the three-dimensional point set
used for MH.

Let $\mathcal X$ be an interval and let
$K_j(x\mid z)$ denote a transition CDF.
Suppose
\[
 \Lambda_j:=
 \sup_{x\in\mathbb R}
 V_{\mathcal X}(K_j(x\mid\cdot))
 <\infty,\qquad j=1,\ldots,T.
\]
Consider the inverse-transform update
\[
 \varphi_j(z,u)
 =\inf\{y\in\mathcal X:K_j(y\mid z)\ge u\},
 \qquad 0<u<1.
\]
Initialize all particles at the same
deterministic state $x_0\in\mathcal X$, with $\nu_0=\delta_{x_0}$,
and use independent
nested uniform scrambles of $(t,m,2)$-nets
with exact first-coordinate stratification,
where $N=b^m$ and $b,t$ are fixed independently of $N$.
Let $\mathcal F_{N,j-1}$ denote the sigma-field generated by the
point-set randomizations through step $j-1$.

For the naturally sorted old array, let
$Z_{N,j-1}(v)$ be the associated step function.
The generalized inverse identity gives the
local event region
\[
 H_{N,j,x}
 =
 \{(v,u)\in[0,1)^2:
   u\le K_j(x\mid Z_{N,j-1}(v))\},
\]
up to the null endpoint $u=0$.
Since $Z_{N,j-1}$ is nondecreasing,
\[
 V_{[0,1)}
 \bigl(K_j(x\mid Z_{N,j-1}(\cdot))\bigr)
 \le\Lambda_j.
\]

The BV-subgraph estimate of
\Cref{lem:threshold-cylinder}, together
with the contribution from the boundary
of the unit square, therefore gives
\[
 \lambda_2((\partial H_{N,j,x})_\varepsilon)
 \le(3\Lambda_j+17)\varepsilon,
 \qquad 0<\varepsilon<1/2,
\]
uniformly in $x$, $N$, and the old array.

{{Define
$e_{N,j}(A_x)=\widehat\nu_{N,j}(A_x)
-\widehat\nu_{N,j-1}(\mathsf P_j\one_{A_x})$ and
$Q_{N,j}g=N^{-1}\sum_{k=0}^{N-1}g(V_{k,j},U_{k,j})$.
Exact sorting-coordinate stratification and the inverse-transform
identity give, almost surely,
\[
 e_{N,j}(A_x)=Q_{N,j}\one_{H_{N,j,x}}-\lambda_2(H_{N,j,x}),
 \qquad
 \E[e_{N,j}(A_x)\mid\mathcal F_{N,j-1}]=0.
\]
The null endpoint $u=0$ is missed almost surely by marginally uniform
points. Conditional on the old array, apply
\Cref{thm:tube-variance} with $D=2$ and
$M=3\Lambda_j+17$ to obtain, for sufficiently large admissible $N$,}}
\[
 \mathbb E_{\mathrm{RQMC}}
 [{{e_{N,j}}}(A_x)^2
  \mid\mathcal F_{N,j-1}]
 \le
 c_{b,t,2}(3\Lambda_j+17)N^{-3/2}.
\]
The conditional BV transfer of
\Cref{lem:bv-transfer}, followed by the
same innovation decomposition and
orthogonality argument used in
\Cref{app:onedim}, then yields
\[
 \mathbb E_{\mathrm{RQMC}}
 [(\widehat\nu_{N,j}(f)-\nu_j(f))^2]
 =
 \mathcal O_j
 \bigl(V_{\mathcal X}(f)^2N^{-3/2}\bigr)
\]
for pointwise $f\in\mathrm{BV}(\mathcal X)$
and fixed $j$.

This recovers the exponent in
\citet[Proposition~6]{lecuyer2008}
under a different point-set assumption:
nested scrambled nets with exact
sorting-coordinate stratification replace
their square-stratified construction.
The argument illustrates the dependence
of the indicator bound on the dimension
of the RQMC point set, without requiring
a separate inversion theorem.

\section{Horizon dependence}
\label{sec:supp-horizon}
We first establish the obstruction to whole-line BV contraction, then
compare this limitation of the bound with the observed horizon dependence.

\subsection{Obstruction to whole-line BV contraction}
We show why the global BV contraction in \Cref{rem:horizon-growth}
fails for whole-line location random-walk MH.

\begin{proposition}[Obstruction to whole-line BV contraction]
\label{prop:no-global-bv-contraction}
For a homogeneous location random-walk MH kernel on $\mathbb R$,
there are no constants $C<\infty$ and $0<\rho<1$ such that
$V_{\mathbb R}(\mathsf P^kf)\le C\rho^kV_{\mathbb R}(f)$ for every
pointwise BV $f$ and every integer $k\ge0$.
\end{proposition}
\begin{proof}
Take $f=\one_{(-\infty,0]}$, with $V(f)=1$. For any fixed $k$, realize the
proposal increments $W_1,\ldots,W_k$ independently of the initial state $z$.
Acceptance only selects which increments are added, so
$|X_k-z|\le S_k:=\sum_{q=1}^k|W_q|<\infty$ almost surely.
For $z>0$, $\mathsf P^kf(z)\le\mathbb P(S_k\ge z)\to0$ as
$z\to\infty$; for $z<0$,
$\mathsf P^kf(z)\ge\mathbb P(S_k<-z)\to1$ as $z\to-\infty$.
Thus $V_{\mathbb R}(\mathsf P^kf)\ge1$ for every finite $k$, contradicting
$C\rho^k\to0$. No proposal moment assumption is used.
\end{proof}
This is an obstruction to that global norm bound, not a variance lower
bound or a refutation of the empirical horizon behavior. For the same
half-line test and $T_0=0$, $T=L$, the function $G_s$ has tail limits
$L-s+1$ and zero. Hence
\[
 \frac1{L^2}\sum_{s=1}^L V_{\mathbb R}(G_s)^2 \ge \frac1{L^2}\sum_{s=1}^L(L-s+1)^2
 =\frac1{L^2}\sum_{q=1}^Lq^2
 =\frac{(L+1)(2L+1)}{6L}.
\]
Thus using a common positive {{$M_s=M$}} in the global BV upper bound does
not yield $L^{-1}$ decay. Local variance control under the laws reached
from $x_0$, weighted variation, or direct covariance estimates would require
additional analysis. Neither ordinary geometric ergodicity from $x_0$ nor
the exact innovation identity supplies a joint $N,T$ rate here.

\subsection{Numerical horizon experiment}
\label{subsec:horizon}

For the random-walk kernels considered here, \Cref{thm:main,rem:horizon-growth}
provide only a worst-case horizon bound. We examine dependence on
\(L=T-T_0\) empirically for the asymmetric Gaussian mixture target and
the CDF {{test}}, using \(N\in\{2^{12},2^{14},2^{16}\}\), \(T_0=2^7\),
  \(T\in\{2^8,2^9,\ldots,2^{16}\}\), \(R_{\mathrm{rep}}=50\),
   and \(x_0=0\).

   {{The Gaussian random-walk proposal has standard deviation
$2.38$, and the CDF threshold is
$a_\pi=F_\pi^{-1}(1/4)\approx-1.4758426711$.
For each fixed $N$, slopes are fitted by ordinary least squares of log sample variance on $\log L$ over the five largest horizons. We obtain 95\% percentile bootstrap intervals by resampling
whole macro-replication estimates independently within
each particle-count, horizon and method combination.}}

\begin{figure}[htbp]
  \centering
  \includegraphics[width=0.85\textwidth]
{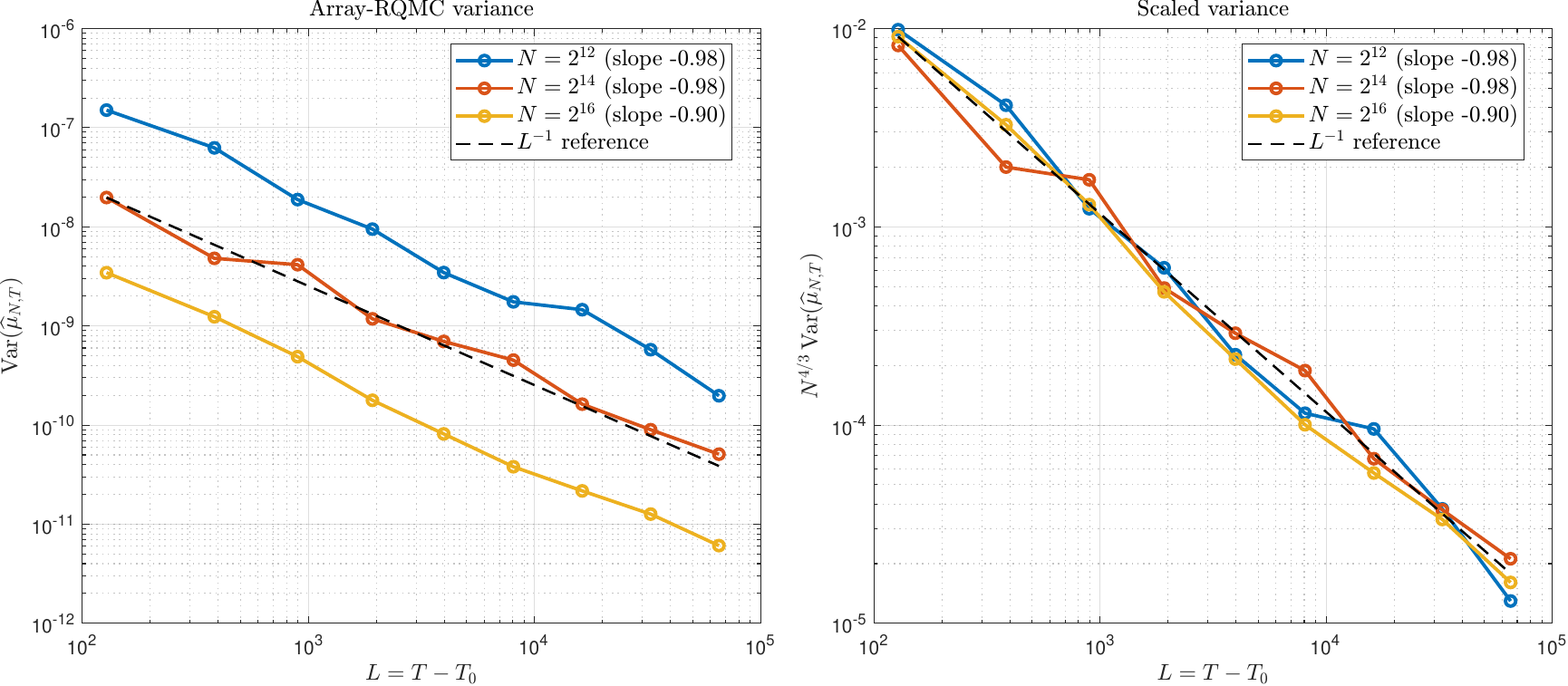}
  \caption{Array-RQMC variance as a function of the averaging length
  \(L=T-T_0\) for the asymmetric Gaussian mixture target and the CDF
  test. Left: replicate sample variance. Right: the same variance scaled
  by \(N^{4/3}\). The dashed lines show the \(L^{-1}\) reference slope.}
  \label{fig:variance-horizon}
\end{figure}

We fit log--log slopes against $L=T-T_0$ using the five largest
horizons, $T=2^{12},\ldots,2^{16}$.
For $N=2^{12},2^{14},2^{16}$, respectively, the fitted slopes are
$-0.975$, $-0.979$, and $-0.898$, with $95\%$ percentile bootstrap
confidence intervals $[-1.122,-0.826]$, $[-1.169,-0.802]$, and
$[-1.051,-0.717]$.
The intervals use 1,000 bootstrap resamples of whole macro-replication
estimates, resampled independently across horizons.
All three intervals contain $-1$, consistent with an approximately
$L^{-1}$ variance decay over the fitted range in
\Cref{fig:variance-horizon}.
The $N^{4/3}$-scaled variance curves in the right panel are of
comparable magnitude and exhibit similar decay across the three
particle counts.
These observations establish neither a joint $(N,T)$ rate nor
a bound on bias relative to the stationary target.

The conservative horizon dependence comes from the global BV estimate
$V_{\mathcal X}(G_s)\le V_{\mathcal X}(f)\mathcal H_s^{T_0,T}$
for the propagated test sum in \eqref{eq:md-propagated}, used in the
exact innovation identity \eqref{eq:innovation-variance}.
The contraction refinement in \Cref{rem:horizon-growth} is unavailable
for whole-line location random-walk MH, as shown by
{{\Cref{prop:no-global-bv-contraction} above}}.
The empirical near-$L^{-1}$ behavior motivates sharper local
innovation-variance bounds, state-law-weighted variation, or direct
covariance estimates.

\FloatBarrier


\paragraph*{Acknowledgement}

The work of Josef Dick was supported in part by a grant of access to OpenAI models through the ChatGPT for Academic Researchers program.

\paragraph*{Declaration of generative AI use}

The authors used ChatGPT 6 Astra and 6.1 Sol for literature searches, exploratory development, and the preparation of portions of the exposition and LaTeX source. The paper has been reviewed using Claude Opus 5.5. All mathematical arguments, calculations, references, and conclusions were independently checked and verified by the authors, who take full responsibility for the contents of the paper.

\bibliographystyle{jcgs-asa}
\bibliography{references_jcgs}


\end{document}